\documentclass[aps,pra,twocolumn,groupaddress,showpacs]{revtex4-1}
\usepackage{amsmath,amssymb,graphicx,stmaryrd}

\usepackage[utf8]{inputenc}
\usepackage[T1]{fontenc}
\usepackage{xcolor}

\IfFileExists{newtxtext.sty}
   {\usepackage{newtxtext,newtxmath}}
   {\IfFileExists{stix.sty}
      {\usepackage{stix}}
      {\IfFileExists{mathptmx.sty}
      {\usepackage{mathptmx}}{} } }

\usepackage{textcomp}

\usepackage{bm}
\usepackage{bbold}

\IfFileExists{siunitx.sty}{\usepackage{booktabs,siunitx}}{}
\pdfoutput=1
\usepackage{color}
\definecolor{LinkColor}{rgb}{0.256,0.439,0.588}
\usepackage{hyperref}
\hypersetup{
   pdfauthor={K. Wierschem and K. S. D. Beach},
   pdftitle={Detection of symmetry-protected topological order in AKLT states by exact evaluation of the strange correlator},
   pdfsubject={Symmetry-protected topological order},
   pdfkeywords={SPT phases,} {topological order,} {AKLT,} {Affleck-Kennedy-Lieb-Tasaki,} {spin systems,} {quantum magnetism,} {strange correlator,} {winding number},
   colorlinks=true,
   citecolor=LinkColor,
   linkcolor=LinkColor,
   urlcolor=LinkColor
   }
\newcommand{\pdoi}[7]{#1, #2, \href{http://dx.doi.org/#7}{ #3 {\bf #4}, #5 (#6)}.}
\newcommand{\purl}[7]{#1, #2, \href{http://#7}{ #3 {\bf #4}, #5 (#6)}.}

\renewcommand{\vec}[1]{\mathbf{#1}}
\newcommand{\bra}[1]{\langle#1\rvert}
\newcommand{\ket}[1]{\lvert#1\rangle}
\newcommand{\bigket}[1]{\bigl\lvert#1\bigr\rangle}
\newcommand{\overlap}[2]{\langle#1\vert#2\rangle}
\newcommand{\expectation}[1]{\langle#1\rangle}

\begin{document}

\title{Detection of symmetry-protected topological order in AKLT states\\ 
by exact evaluation of the strange correlator}

\author{K. Wierschem}
\email[Electronic mail:\ ]{keola@olemiss.edu}
\affiliation{Department of Physics and Astronomy, The University of Mississippi, University, Mississippi 38677, USA}

\author{K. S. D. Beach}
\email[Electronic mail:\ ]{kbeach@olemiss.edu}
\affiliation{Department of Physics and Astronomy, The University of Mississippi, University, Mississippi 38677, USA}

\date{July 1, 2016}

\begin{abstract}
The strange correlator [Phys. Rev. Lett. {\bf 112}, 247202 (2014)] has been proposed as a measure of symmetry protected topological order in one- and two-dimensional systems. It takes the form of a spin-spin correlation function, computed as a mixed overlap between the state of interest and a trivial local product state. We demonstrate that it can be computed exactly (asymptotically, in the Monte Carlo sense) for various Affleck-Kennedy-Lieb-Tasaki states by direct evaluation of the wave function within the valence bond loop gas framework. We present results for lattices with chain, square, honeycomb, cube, diamond, and hyperhoneycomb geometries. In each case, the spin quantum number $S$ is varied such that $2S$ (the number of valence bonds emerging from each site) achieves various integer multiples of the lattice coordination number. We introduce the concept of strange correlator loop winding number and point to its utility in testing for the presence of symmetry protected topological order.
\end{abstract}


\maketitle

\section{ \label{SEC:introduction} Introduction }

The classification of topological phases of matter is increasingly well understood~\cite{Senthil-ARCMP-15}. We now know that, in addition to states with {\it intrinsic} topological order~\cite{Wen-IJMP-90}---which cannot be smoothly connected to any local product state~\cite{Chen-PRB-10}---there are those that exhibit a weaker form of topological order. These so-called {\it symmetry protected topological} (SPT) states~\cite{Gu-PRB-09,Pollmann-PRB-12a} are also incompatible with any local-product-state description, but only insofar as a special protecting symmetry remains unbroken. 

SPT states are gapped and featureless in the bulk, and they are short-range entangled. They share few of the exotic features of their intrinsically ordered cousins. For instance, they do not support fractional excitations~\cite{Laughlin-PRL-83}, nor do they possess topological ground-state degeneracy~\cite{Niu-PRB-90} (i.e., they have unique ground states on closed manifolds). Nonetheless, nontrivial SPT states do possess interesting boundary modes, which in low-dimensional systems must either be gapless or spontaneously break a symmetry. (Beginning in three spatial dimensions, topologically degenerate boundary modes offer a third possibility~\cite{Vishwanath-PRX-13}.) In general, the wave function at the boundary admits projective representations of the protecting symmetries. The Haldane chain is a well-known example of an SPT state in one dimension, where the spin-1 degrees of freedom in the bulk fractionalize into spin-1/2 at the boundary~\cite{Polizzi-PRB-98}; there, the projective representations are SU(2) in the case of SO(3) symmetry and ${\cal T}^2=-1$ in the case of time reversal symmetry.

The concept of SPT phases was first developed in the context of noninteracting fermions, and an exhaustive classification~\cite{Schnyder-PRB-08,Kitaev-AIP-09} was soon worked out following the discovery of topological insulators. In the presence of interactions, however, the story is more complicated. For example, in the case of spinless fermions in one dimension with time-reversal invariance and particle-number conservation, interactions reduce the group structure from ${\mathbb Z}$ down to ${\mathbb Z}_8$~\cite{Fidkowski-PRB-10,Turner-PRB-11}; for bosons, the group structure must enlarge, since noninteracting bosons ultimately condense into a single phase. There is ongoing work to classify SPT phases for general interacting quantum systems. For fermions, there is a proposal to apply group supercohomology theory~\cite{Gu-PRB-14}. Classification of interacting bosonic systems has been attempted using group cohomology~\cite{Chen-PRB-13} and nonlinear $\sigma$ models augmented by topological $\theta$ terms~\cite{Bi-PRB-15}.

Much progress has been made already in the special case of one spatial dimension, where gapped SPT phases can be described by matrix product states~\cite{Schuch-PRB-11,Pollmann-PRB-12b} and their characterization given by a generalization of string order. But in two or more dimensions, string order is ill-defined, and such a classification scheme is not possible. Some promising approaches include universal signatures in the entanglement spectrum~\cite{Li-PRL-08} and the braiding statistics of topological excitations in the corresponding ``gauged'' intrinsic topological state~\cite{Levin-PRB-12}. Recent proposals extend to the geometric properties of reduced density matrices~\cite{Chen-PRA-16} and the analysis of modular matrices~\cite{Huang-arXiv-15}. However, these approaches are computationally costly and impractical outside of exact diagonalization, density matrix renormalization group, matrix product state, and tensor network methods.

An intriguing alternative has been suggested by You {\it et al.}, who introduced a ``strange correlator'' that effectively maps spatiotemporal correlations at the physical boundary of an SPT phase to spatial correlations at a temporal boundary with a trivial symmetric product state~\cite{You-PRL-14}. Because the modes at the physical boundary are either gapless or degenerate, in one and two dimensions they must be either critical or ordered (because of spontaneous symmetry breaking at the boundary). Thus, the strange correlator will show exponential decay for trivial symmetric states but long-range or power-law behavior for nontrivial SPT states. This understanding has been confirmed in studies of Affleck-Kennedy-Lieb-Tasaki (AKLT) states in one and two dimensions~\cite{You-PRL-14}, spin-1 Heisenberg chains and ladders~\cite{Wierschem-PRB-14}, quasi-one-dimensional arrays of Haldane chains~\cite{Wierschem-PRL-14}, and quantum spin Hall states in the Kane-Mele-Hubbard model~\cite{Wu-PRB-15}.

The basic idea of the strange correlator is to map the imaginary-time Green's function of a spatial boundary onto the equal-time Green's function of a space-time boundary~\cite{You-PRL-14}. As long as there exists an emergent Lorentz symmetry in the low-energy effective field theory of the SPT phase (i.e., the dynamical critical exponent $z=1$), then this mapping is in principle exact. Since the boundary modes of an SPT are either gapless or degenerate, in one and two dimensions this implies that the Green's function must show either long-range order (LRO) or quasi-long-range order (QLRO). In three dimensions, the possibility of a topologically ordered boundary mode introduces the possibility of short-range order (SRO). Thus, in one and two dimensions, LRO or QLRO signal a nontrivial SPT phase, while SRO signals a symmetric product state. In three dimensions, LRO and QLRO still signal a nontrivial SPT phase, but SRO no longer implies a symmetric product state unless it can be shown that the boundary is topologically trivial.

In this work, we demonstrate how to measure the strange correlator within a generalized spin-$S$ valence bond formalism and apply this method to study AKLT states in one, two, and three dimensions. The AKLT states are exact valence bond solid states with well-known properties, and we employ them as a testbed for methods to characterize SPT states. In one dimension our approach provides an intuitive and easy-to-visualize picture of why the strange correlator yields long-range correlations in spin-$S$ AKLT chains for odd values of $S$. It also makes clear a connection to the concept of valence bond winding number sectors, as well as to the conventional string order in one-dimensional systems. In two dimensions the strange correlator becomes quasi-long-range, yet we show that the corresponding winding number fraction remains finite. This is reminiscent of the Kosterlitz-Thouless phase, where slow twisting of the superfluid order parameter prevents Bose-Einstein condensation at finite temperature, yet there still exists a finite superfluid fraction due to vortex confinement. Finally, in three dimensions we show the strange correlator is long-range with a nonzero winding number fraction.

The remainder of the text proceeds as follows. In Sec.~\ref{SEC:Method} we introduce AKLT states as the exact ground states of a class of projective Hamiltonians. We also present a generalized valence bond formalism for spin-$S$ degrees of freedom composed of symmetrized combinations of $2S$ spin-1/2 particles and show how to measure correlation functions, conventional and strange, within this formalism. In Sec.~\ref{SEC:Results}, we present our results, which include a numerical evaluation of these correlation functions (for various  lattices and spin values) and a comparison to some exact results in one dimension. In Sec.~\ref{SEC:energy} we show that the energy excitation gap of AKLT states can be estimated by promoting one valence bond singlet to a triplet (in the spirit of Feynman's single-mode approximation) and give some estimates for states with $S\le2$. Finally, in Sec.~\ref{SEC:Conclusions}, we summarize our main results and discuss the outlook for future work.

\begin{figure}
\begin{center}
\includegraphics{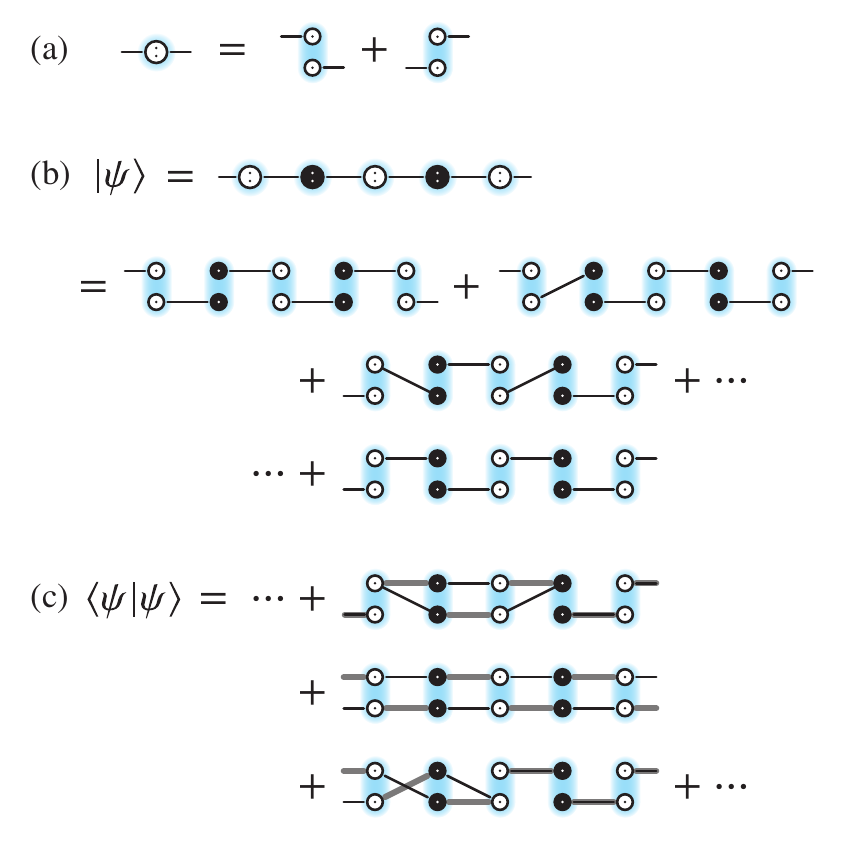}
\end{center}
\caption{\label{FIG:spin-half-symmetrization} 
(a) Each $S=1$ object can be expressed as two spin-half degrees of freedom, fully symmetrized. (b) The AKLT state on the linear chain is the state with exactly one short bond connecting each pair of nearest-neighbor spins. Here, all bonds represent singlet pairs between sites in opposite sublattices (denoted by open and filled circles). When the two spin-half flavors on each site are resolved, the state takes the form of a superposition of $2^N$ nonorthogonal valence bond configurations. (c) The overlap of the AKLT state with itself produces a gas of closed loops. This is the framework in which all expectation values are computed, as per Ref.~\onlinecite{Beach-NPB-06}.
}
\end{figure}

\begin{figure}
\begin{center}
\includegraphics{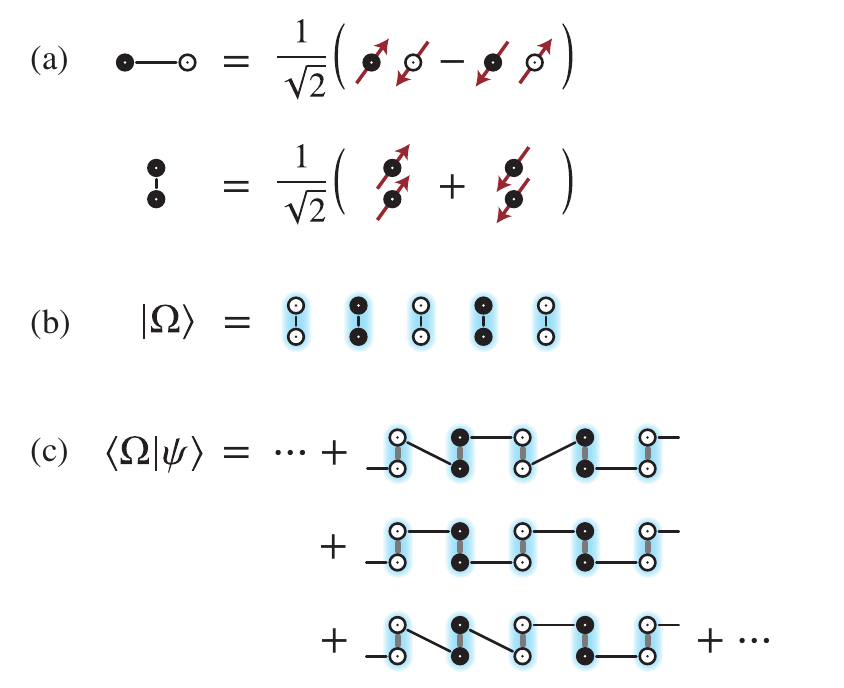}
\end{center}
\caption{\label{FIG:strange-correlator} 
(a) Our convention is that a solid line connecting spins in opposite sublattices represents a singlet bond; a line connecting spins in the same sublattice represents a triplet. For our purposes, we have selected $\ket{t} = \bigl(\ket{S=1; S^z = +1} + \ket{S=1; S^z = -1}\bigr)/\sqrt{2}$, the symmetric triplet of aligned spin pairs. (b) The reference state is a tensor product of such triplets on each site, $\ket{\Omega} = \otimes_i \ket{t}_i$. (c) The overlap of the reference state with the AKLT state from Fig.~\ref{FIG:spin-half-symmetrization}(b) is depicted. A special property of the 1D system is that every term in this mixed overlap involves a single, system-spanning loop whenever $\ket{\psi}$ is a nontrivial SPT state.
}
\end{figure}

\begin{figure}
\begin{center}
\includegraphics{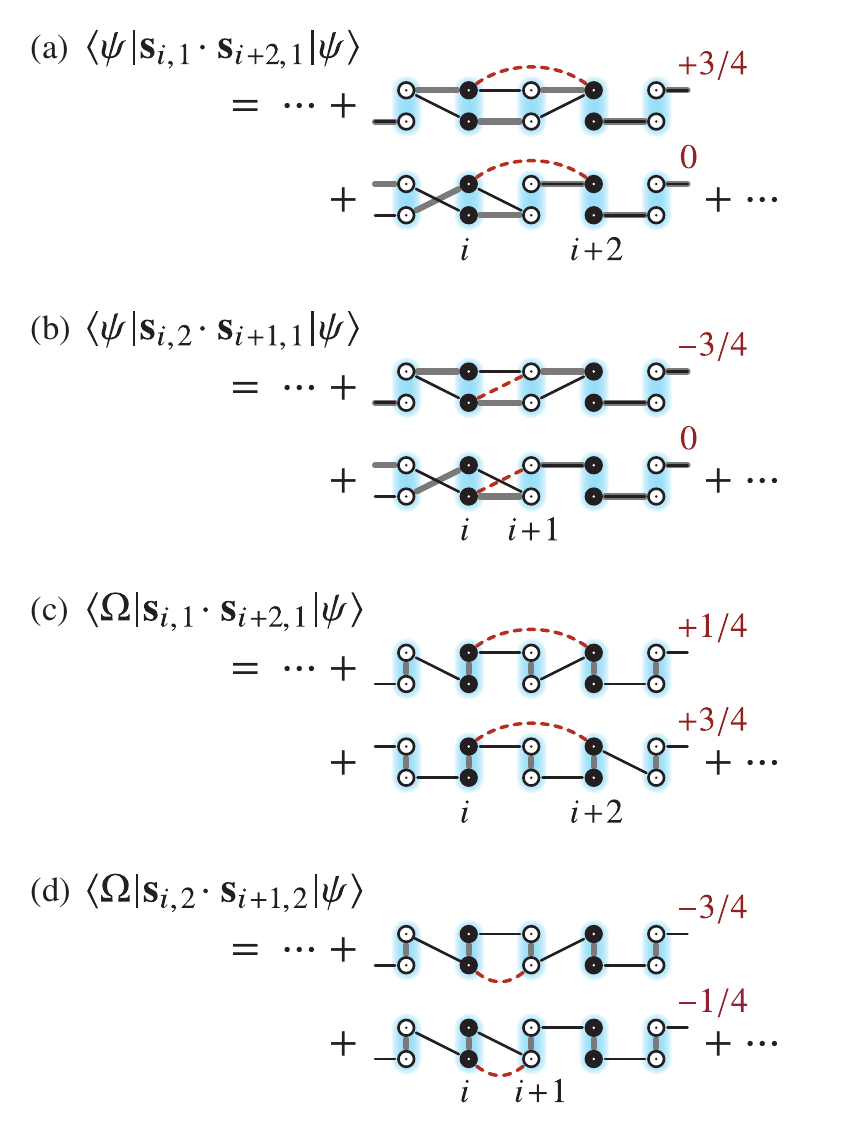}
\end{center}
\caption{\label{FIG:expectation-values} 
Spin correlations between the constituent spin-halfs obey the usual rules of the valence bond basis: (a) $+3/4$ for spin-halfs in the same sublattice and (b) $-3/4$ for spin-halfs in opposite sublattices, provided they reside in the same loop of the overlap graph; zero otherwise. For mixed estimators involving the triplet product state, the same rules apply so long as the {\it loop spacing} obeys the expected $ABAB$ pattern; otherwise the value is a factor of 3 lower: (c) for spin-halfs in the same sublattice, $+1/4$ for {\it odd} loop spacing and $+3/4$ for the usual even loop spacing, and (d) in opposite sublattices, $-3/4$ for the usual odd loop spacing and $-1/4$ for {\it even} loop spacing.
}
\end{figure}

\section{\label{SEC:Method} Model and Methods}

The AKLT~\cite{Affleck-PRL-87,Affleck-CMP-88} states are prototypical examples of SPT states in low dimension. (They have recently gained new attention for their possible value in measurement-based quantum computation schemes~\cite{Wei-PRL-11}.) The best-known example in one dimension is the Haldane chain, which is smoothly connected to an exact valence bond solid state~\cite{Affleck-PRL-87}. By expressing its Hamiltonian as a sum of projection operators on nearest-neighbor bonds, one can prove that there is a unique ground state (on the periodic chain) with exponentially decaying correlations and a gap to excitations~\cite{Affleck-CMP-88}. Exact results for the bilinear spin correlation function of the spin-$S$ AKLT chain have been derived on the basis of an exact mapping from a $D$-dimensional AKLT state to a $D$-dimensional model of interacting classical O(3) spins at finite temperature~\cite{Arovas-PRL-88}. This mapping holds in any dimension and implies, via the Mermin-Wagner theorem~\cite{Mermin-PRL-66}, that all AKLT states in one and two dimensions are magnetically disordered. In 3D, magnetic order is not forbidden and its presence is specific to the choice of lattice and spin value. Ordering typically occurs above some threshold in the {\it multiplicity} of the state (viz., the integer $M = 2S/z$, where $z$ is the coordination number of the lattice). For instance, Monte Carlo simulations have shown that AKLT states on the cubic lattice always exhibit long-range antiferromagnetic order, whereas those on the diamond lattice are ordered only if there is more than one singlet bond per nearest-neighbor link~\cite{Parameswaran-PRB-09}.

The AKLT states are the ground states of Hamiltonians of the form
\begin{equation} \label{EQ:AKLT_Hamiltonian}
\hat{H} = \sum_{\langle i,j \rangle}\,\,\sum_{J=2S-M+1}^{2S}A^{}_{J}P_{J}^{2S}(i,j).
\end{equation}
Here, the first sum is over nearest-neighbor pairs of sites, each hosting a local spin-$S$ degree of freedom. The second sum is over spin sectors, with $M$ being the multiplicity of the AKLT state. The projector $P_{J}^{2S}(i,j)$ projects onto the total spin $J$ sector of the combined spin space $2S$ between sites $i$ and $j$, and the $A_{J}$ are interaction strengths of the projectors. The AKLT state is the ground state as long as the interactions satisfy $A_{J}>0$, so in practice their precise values are arbitrary (though they do set the energy scale for excitations). The spin projectors $P_{J}^{2S}(i,j)$ can be given an explicit form by recognizing that their effect is to eliminate all sectors in which the angular momentum sum does not produce a net spin of quantum number $J$,
\begin{equation} \label{EQ:Explicit_form_of_the_projector}
P_{J}^{2S}(i,j) = \prod_{\substack{J'=0\\(J'\neq J)}}^{2S}\frac{\bigl(\vec{S}_{i}+\vec{S}_{j}\bigr)^{2}-J'\bigl(J'+1\bigr)}{J\bigl(J+1\bigr)-J'\bigl(J'+1\bigr)}.
\end{equation}
The projectors can be recast in terms of Heisenberg spin interactions via the identity $\bigl(\vec{S}_{i}+\vec{S}_{j}\bigr)^{2}=2S(S+1)+2\vec{S}_{i}\cdot\vec{S}_{j}$. Note that the particular terms in the denominator of Eq.~\eqref{EQ:Explicit_form_of_the_projector} ensure that the completeness relation $\sum_{J=0}^{2S}P_{J}^{2S}(i,j)=1$ is satisfied. (Projective constructions such as this are commonplace; e.g., see Eq.~(2) of Ref.~\onlinecite{Parameswaran-PRB-09} or Eq.~(8) of Ref.~\onlinecite{Beach-PRB-09b}. Table~\ref{TAB:AKLT-coefficients} gives the projectors as a power series in the Heisenberg spin interactions for $S\le3$.) For the case of multiplicity $M=1$, the AKLT Hamiltonian takes a rather simple form with $P_{2S}^{2S}(i,j)$ as the only contributing element. For the $S=1$ case, we obtain the celebrated AKLT point of the bilinear-biquadratic chain, namely $P_{2}^{2}(i,j)=\frac{1}{3}+\frac{1}{2}(\vec{S}_{i}\cdot\vec{S}_{j})+\frac{1}{6}(\vec{S}_{i}\cdot\vec{S}_{j})^{2}$, while for $S=3/2$, $S=2$ and $S=3$ we recover the standard ($M=1$) AKLT Hamiltonians on lattices with coordination $z=3$ (honeycomb, hyperhoneycomb), $z=4$ (square, diamond), and $z=6$ (cubic), respectively.

\begin{table}
\caption{The projectors $P^{2S}_{J}(i,j)$ are expressed as a series expansion
in the spin interactions $(\vec{S}_{i}\cdot\vec{S}_{j})^{m}$.
The expansion coefficients at order $m$ are given for select values of $0 \le J \le S$.}
\begin{tabular}{ c c c c c c c c }
\hline \hline
& \multicolumn{7}{c}{$0 \le m \le 2S$}\\ \cline{2-8}
$P^{2S}_J$ & 0 & 1 & 2 & 3 & 4 & 5 & 6\\[+0.05cm]
\hline
$P^{2}_{0}$ & $-\frac{1}{3}$ & $\phantom{+}0$ & $\phantom{+}\frac{1}{3}$ & & & & \\[-0.05cm]
$P^{2}_{1}$ & $\phantom{+}1$ & $-\frac{1}{2}$ & $-\frac{1}{2}$ & & & & \\[-0.05cm]
$P^{2}_{2}$ & $\phantom{+}\frac{1}{3}$ & $\phantom{+}\frac{1}{2}$ & $\phantom{+}\frac{1}{6}$ & & & & \\[0.1cm]

$P^{3}_{0}$ & $\phantom{+}\frac{33}{128}$ & $\phantom{+}\frac{31}{96}$ & $-\frac{5}{72}$ & $-\frac{1}{18}$ & & & \\[-0.05cm]
$P^{3}_{1}$ & $-\frac{81}{128}$ & $-\frac{117}{160}$ & $\phantom{+}\frac{9}{40}$ & $\phantom{+}\frac{1}{10}$ & & & \\[-0.05cm]
$P^{3}_{2}$ & $\phantom{+}\frac{165}{128}$ & $\phantom{+}\frac{23}{96}$ & $-\frac{17}{72}$ & $-\frac{1}{18}$ & & & \\[-0.05cm]
$P^{3}_{3}$ & $\phantom{+}\frac{11}{128}$ & $\phantom{+}\frac{27}{160}$ & $\phantom{+}\frac{29}{360}$ & $\phantom{+}\frac{1}{90}$ & & & \\[0.1cm]

$P^{4}_{0}$ & $\phantom{+}0$ & $-\frac{1}{3}$ & $-\frac{17}{180}$ & $\phantom{+}\frac{1}{45}$ & $\phantom{+}\frac{1}{180}$ & & \\[-0.05cm]
$P^{4}_{1}$ & $\phantom{+}0$ & $\phantom{+}\frac{4}{5}$ & $\phantom{+}\frac{1}{5}$ & $-\frac{1}{18}$ & $-\frac{1}{90}$ & & \\[-0.05cm]
$P^{4}_{2}$ & $\phantom{+}0$ & $-\frac{20}{21}$ & $-\frac{1}{9}$ & $\phantom{+}\frac{1}{18}$ & $\phantom{+}\frac{1}{126}$ & & \\[-0.05cm]
$P^{4}_{3}$ & $\phantom{+}1$ & $\phantom{+}\frac{9}{20}$ & $-\frac{7}{360}$ & $-\frac{1}{36}$ & $-\frac{1}{360}$ & & \\[-0.05cm]
$P^{4}_{4}$ & $\phantom{+}0$ & $\phantom{+}\frac{1}{28}$ & $\phantom{+}\frac{1}{40}$ & $\phantom{+}\frac{1}{180}$ & $\phantom{+}\frac{1}{2520}$ & & \\[0.1cm]

$P^{6}_{6}$ & $-\frac{1}{175}$ & $-\frac{7}{1980}$ & $\phantom{+}\frac{1}{10395}$ & $\phantom{+}\frac{821}{2494800}$ & $\phantom{+}\frac{487}{7484400}$ & $\frac{37}{7484400}$ & $\frac{1}{7484400}$\\[+0.05cm]
\hline \hline
\end{tabular}
\label{TAB:AKLT-coefficients}
\end{table}

The AKLT states are fixed-point representations of valence bond solids. To see this, it is instructive to decompose the spin operator $\vec{S}_{i}$ on a lattice site into $2S$ spin-1/2 components~\cite{Affleck-PRL-87,Affleck-CMP-88}. In this picture, AKLT states are formed by combining the spin-1/2 components on adjacent lattice sites into a spin singlet in such a way that the number of singlets across each nearest-neighbor link is a 
constant (the multiplicity). Then the spin-1/2 components residing at each site are projected onto the total spin-$S$ sector.

For concreteness, we proceed by expressing each spin-$S$ operator ($\vec{S}_i$)
as a sum 
\begin{equation}
\vec{S}_i = \sum_{\alpha=1}^{2S} \vec{s}_{i,\alpha}
\end{equation}
of spin-half ``flavors'' ($\vec{s}_{i,1}, \vec{s}_{i,2}, \ldots, \vec{s}_{i,2S}$)
and forcing all $2S$ of them into ferromagnetic alignment. Since
$\hat{Q}_{i,\alpha;j,\beta} = 1/4 + \vec{s}_{i,\alpha}\cdot\vec{s}_{j,\beta}$
functions as a bond swap operator when the spin flavors reside on the same sublattice
[see Refs.~\onlinecite{Liang-PRL-88} and \onlinecite{Lou-PRB-07}; cf.\ Eqs.~(4) and (13) in Ref.~\onlinecite{Beach-PRB-09a}],
maximal ferromagnetic alignment is equivalent to perfect symmetrization
of the $2S$ spin flavors:
\begin{equation}
\begin{split}
\ket{\psi\{\vec{S}_i\}} &\sim 
\lim_{g \to \infty} \exp\biggl(g \sum_i \sum_{\alpha < \beta}
\vec{s}_{i,\alpha}\cdot\vec{s}_{i,\beta}\biggr)\\
&\qquad\qquad\qquad \times \ket{\psi\{\vec{s}_{i,1},\vec{s}_{i,2},\cdots,\vec{s}_{i,2S}\}}\\
&\sim \frac{1}{(2S)!}\sum_\pi\ket{\psi\{\vec{s}_{i,\pi(1)},\vec{s}_{i,\pi(2)},\cdots,\vec{s}_{i,\pi(2S)}\}},
\end{split}
\end{equation}
where $g$ represents a ferromagnetic coupling factor between the constituent spin-half degrees of freedom, and $\pi$ labels the $(2S)!$ possible permutations of the flavor indices (which is what we sample in our Monte Carlo simulations). AKLT states are obtained by applying the symmetrization above to a fixed pattern of valence bonds; namely, $M$ valence bonds between spin flavors of nearest neighbors. As an example, in Fig.~\ref{FIG:spin-half-symmetrization} we illustrate the $M=1$ AKLT state on a spin-1 chain.

Within this representation, the spin correlations of the spin-$S$ entities can be decomposed into the constituent spin correlations of the spin-half flavors,
\begin{equation}\label{EQ:spin-correlations}
\begin{split}
\bra{v} \vec{S}_i\cdot\vec{S}_j \ket{v'}
&= \sum_{\alpha=1}^{2S}\sum_{\beta=1}^{2S}\bra{v}\vec{s}_{i,\alpha}\cdot\vec{s}_{j,\beta}\ket{v'}.
\end{split}
\end{equation}
Here, $v$ and $v'$ denote bipartite~\cite{Beach-NPB-06} spin-$S$ valence bond configurations, each with $SN$ valence bonds connecting $N/2$ sites in the A sublattice to $N/2$ sites in the B sublattice. Thus, we can evaluate the flavor correlations $\bra{v}\vec{s}_{i,\alpha}\cdot\vec{s}_{j,\beta}\ket{v'}$ using a standard result for the valence bond basis~\cite{Beach-NPB-06},
\begin{equation}\label{EQ:flavor-correlations}
\frac{\bra{v}\vec{s}_{i,\alpha}\cdot\vec{s}_{j,\beta}\ket{v'}}{\overlap{v}{v'}} = \frac{3}{4}\epsilon_{i,j}\delta_{i,\alpha \leftrightarrow j,\beta}.\\
\end{equation}
Here, $\epsilon_{i,j}$ takes the value $+1$ or $-1$, depending on whether or not the sites $i$ and $j$ share a common sublattice label. 
The quantity $\delta_{i,\alpha \leftrightarrow j,\beta} = 0,1$ is a $\delta$ function that triggers when spin flavors $i,\alpha$ and $j,\beta$ reside in the same loop. The possible outcomes for Eq.~\eqref{EQ:flavor-correlations}, in the case of a spin-1 chain, are illustrated in Figs.~\ref{FIG:expectation-values}(a) and \ref{FIG:expectation-values}(b).

Another useful result is the overlap of two valence bond configurations, given by $\overlap{v}{v'}=2^{N_{\ell}-SN}$. Here, $N_{\ell}$ is the number of loops in the overlap graph, which is formed by laying configuration $v$ atop configuration $v'$. The maximum possible number of loops in an overlap graph is $SN$, which only occurs when two valence bond configurations are identical (and hence the overlap is unity).

As we emphasized previously, symmetrization~\cite{Affleck-PRL-87} between the $2S$ spin-half flavors on a given site is equivalent to a swap of the end points of two valence bonds. With respect to Monte Carlo evaluation, this symmetrization is the only kind of dynamical fluctuation one needs to account for, since the AKLT states are realizations of otherwise static valence bond patterns. 

Accordingly, we have implemented the following Monte Carlo update scheme. We sweep through the lattice, choosing each spin-$S$ degree of freedom in turn. For each spin, we  select at random a pair of spin-half flavors and propose a new configuration with the end points of these two flavors swapped (see Fig.~\ref{FIG:loop-rules-parity} in the Appendix for an illustration). The new configuration is accepted according to the ratio of the Monte Carlo sampling weight,
\begin{equation}
\frac{W(v,v')^\text{new}}{W(v,v')^\text{old}} = \frac{2^{N_{\ell}^\text{new}-SN}}{2^{N_{\ell}^\text{old}-SN}} = 2^{\Delta N_\ell},
\end{equation}
which depends only on the change in the loop count, $\Delta N_\ell = N_{\ell}^\text{new} - N_{\ell}^\text{old}$.
Although it is sometimes advantageous to perform a ``walk'' through the new configuration to determine the change in the number of loops in the overlap graph~\cite{Beach-PRB-09b,Zhang-PRB-13}, this procedure can be slow, especially if a significant fraction of the loops are large. A fast update scheme exists whereby the change in the number of loops can be effectively sampled by considering the combined bond-spin space, as first discussed by Sandvik and Evertz~\cite{Sandvik-PRB-10}. We discuss our own implementation of this protocol in the Appendix.

Let us define ``normal'' and ``strange'' versions of the bilinear spin correlation function, where we use an overbar to denote mixed estimators such as the strange correlator:
\begin{equation}\label{EQ:bilinear}
C(\vec{r}) = \frac{\bra{\psi}\vec{S}_{0}\cdot\vec{S}_{\vec{r}}\ket{\psi}}{\overlap{\psi}{\psi}},\quad
\bar{C}(\vec{r}) = \frac{\bra{R}\vec{S}_{0}\cdot\vec{S}_{\vec{r}}\ket{\psi}}{\overlap{R}{\psi}}.
\end{equation}
Here, $\ket{\psi}$ is the AKLT state and $\ket{R}$ is a reference state used in defining the strange correlator. The reference state $\ket{R}$ should be a local product state with all the symmetries of $\ket{\psi}$. One choice for $\ket{R}$ is thus a dimer product state with $S$ valence bonds making up each dimer. Such a state falls within the singlet sector and mixed estimator $\bar{C}(r)$ can be sampled in the usual valence bond basis. However, while this choice clearly has all the spin symmetries, it breaks translational symmetry. Instead, we follow You {\it et al.}~\cite{You-PRL-14} in choosing an on-site product state with zero spin projection along one axis, $\ket{\Omega}=\prod_{i}\ket{S_{i}^{y}=0}$ (we use the $y$ axis in place of the $z$ axis for reasons to be explained in the Appendix). This state can be represented in the valence bond basis by pairing up the $2S$ spin flavors on a site into $S$ spin-zero triplets (see Fig.~\ref{FIG:strange-correlator} for an illustration of $\ket{\Omega}$ and $\overlap{\Omega}{\psi}$ for the $S=1$ chain).
We have given an explicit formula for the spin correlations of two valence bond configurations in Eq.~\eqref{EQ:spin-correlations}. This formula is valid for any bipartite valence bond configurations in the total singlet sector. To compute the strange correlator using $\ket{\Omega}$, however, we must make use of on-site ``internal triplet'' states that require new rules. As before, we can decompose the spin correlations $\vec{S}_{i}\cdot\vec{S}_{j}$ into the flavor correlations $\vec{s}_{i,\alpha}\cdot\vec{s}_{j,\beta}$. However, we must now evaluate these correlations in a mixed estimator between a valence bond configuration in the singlet sector $\ket{v}$ and one in the triplet sector $\ket{t}$. This leads to the following formula:
\begin{equation}\label{EQ:strange-correlations}
\frac{\bra{v}\vec{s}_{i,\alpha}\cdot\vec{s}_{j,\beta}\ket{t}}{\overlap{v}{t}} = \biggl(\frac{1}{2}\epsilon_{i,j}+\frac{1}{4}\lambda_{i,\alpha \leftrightarrow j,\beta}\biggr)\delta_{i,\alpha \leftrightarrow j,\beta},
\end{equation}
where $\lambda_{i,\alpha \leftrightarrow j,\beta}$ is the ``loop spacing'' between $i,\alpha$ and $j,\beta$, defined as $+1$ or $-1$ for even and odd distances between flavors in the same loop (and zero otherwise). Note that the usual valence bond formula is recovered by setting $\lambda_{i,\alpha \leftrightarrow j,\beta}=\epsilon_{i,j}$. The modified rules that lead to this formula are derived in the Appendix and illustrated in Figs.~\ref{FIG:expectation-values}(c) and \ref{FIG:expectation-values}(d).

Note that a similar mixed correlator has been developed to determine unambiguously the valence bond occupation number in bipartite valence bond states~\cite{Alet-PRB-10,Schwandt-PRB-14}. In that application, however, the N\'{e}el state is used as the reference state. In this work, the N\'{e}el state is inappropriate, since it breaks all the protecting symmetries of the the AKLT states---viz., time-reversal invariance and the SO(3) and dihedral ${\mathbb Z}_{2}\times {\mathbb Z}_{2}$ symmetries.

In addition to the spin correlation functions described above, we can also define a winding fraction, analogous to the spin stiffness or superfluid fraction. Here, instead of using the winding number of a conserved charge, we utilize the valence bond winding number, defined as
\begin{equation}
\expectation{W_{\alpha}^{2}} = \Bigl\langle\sum_{\ell}\biggl(W_{\alpha}^{\left(\ell\right)}\biggr)^{2}\Bigr\rangle,
\end{equation}
where the sum is over all loops $\ell$ of the overlap graph and $W_{\alpha}^{(\ell)}$ is the winding number of loop $\ell$ in the $\alpha$ direction. The normal and strange winding fractions are then
\begin{equation}
\label{EQ:winding-fraction}
\rho = \frac{\sum_{\alpha=1}^{d}\expectation{W_{\alpha}^{2}}}{dSN},
\quad
\bar{\rho} = \frac{\sum_{\alpha=1}^{d}\expectation{\bar{W}_{\alpha}^{2}}}{dSN}.
\end{equation}
If $\rho$ remains constant as $N\rightarrow\infty$, this is a sign of long-range or quasi-long-range bulk correlations. Similarly, if $\bar{\rho}$ remains constant as $N\rightarrow\infty$, then by the mapping of the strange correlator onto the boundary modes, this indicates long-range or quasi-long-range boundary correlations, a sign of nontrivial SPT order.

Another useful estimator is the total staggered magnetization, defined as $M^{2}=\sum_{i,j}\epsilon_{i,j}\vec{S}_{i}\cdot\vec{S}_{j}$. In the valence bond framework, it can be shown~\cite{Beach-NPB-06} that $M^{2}$ is related to the average squared length of loops in the loop gas picture. The exact relations are
\begin{equation}
\label{EQ:staggered-magnetization}
M^{2} = \frac{3}{4} \sum_{\ell} \Bigl\langle L_{\ell}^{2} \Bigr\rangle,
\quad
\bar{M}^{2} = \frac{1}{2} \sum_{\ell} \Bigl\langle L_{\ell}^{2} \Bigr\rangle,
\end{equation}
where $L_{\ell}$ is the length of loop $\ell$ and the different prefactors stem from the difference between Eqs.~\eqref{EQ:flavor-correlations} and ~\eqref{EQ:strange-correlations} .

\section{\label{SEC:Results} Results }

\subsection{ \label{SSEC:1D} Linear chains }

Historically, the AKLT chain was the first robust example of an integer spin-$S$ antiferromagnet with short-range correlations and a gap to all excited states~\cite{Affleck-PRL-87}, in confirmation with Haldane's conjecture~\cite{Haldane-PLA-83,Haldane-PRL-83}.

The bilinear spin correlations in spin-$S$ AKLT chains have been calculated exactly by Arovas {\it et al.}~\cite{Arovas-PRL-88}:
\begin{equation}\label{EQ:chain-normal}
C(r) = (-1)^{r}(S+1)^{2}\biggl(\frac{S}{S+2}\biggr)^{r}.
\end{equation}
We find this equation also to be valid for the strange correlator when $S$ is even. When $S$ is odd, however, the strange correlator appears to obey
\begin{equation}\label{EQ:chain-strange}
\bar{C}(r) = (-1)^{r}\frac{(S+1)^{2}}{2}\biggl[1+\biggl(\frac{S-1}{S+3}\biggr)^{r}\biggr].
\end{equation}

\begin{figure}
\begin{center}
\includegraphics[clip,trim=0cm 0cm 0cm 0cm,width=\linewidth]{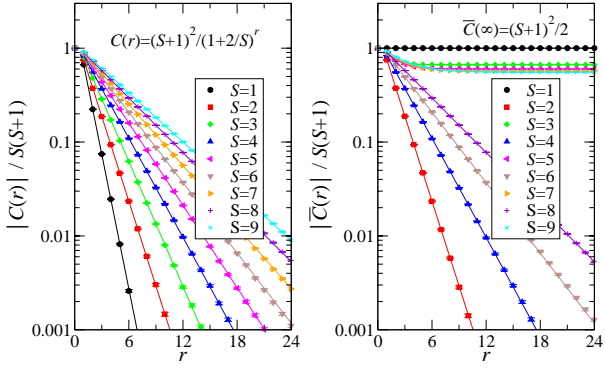}
\end{center}
\caption{\label{FIG:fig-chain} 
Normal (left) and strange (right) spin correlation functions for spin-$S$ AKLT chains. Lines are exact forms, while data points are sampled according to the Monte Carlo scheme described in the main text. The normal spin correlation functions are given by Eq.~\eqref{EQ:chain-normal}, which is also the correct form for strange correlations when $S$ is even. For odd $S$, the empirical form is given by Eq.~\eqref{EQ:chain-strange}.}
\end{figure}

These scaling forms are compared in Fig.~\ref{FIG:fig-chain} to results from our MC sampling scheme for spin-$S$ AKLT chains with spin values ranging from $S=1$ to $S=9$. For $S=1$, we replicate the result of You {\it et al.}~\cite{You-PRL-14}, $\bar{C}(\infty)=2$, while for generic odd integral $S$ we find $\bar{C}(\infty)=(S+1)^{2}/2$. For $S$ an even integer, we find $\bar{C}(\infty)=0$. Thus, the strange correlator correctly captures the even-odd effect in this system: for $S$ even, the edge spins are integral and the bulk state is smoothly connected to a product state, while for $S$ odd, the edge spins are fractional and the bulk cannot be smoothly connected to a product state as long as the symmetry is preserved.

\begin{figure}
\begin{center}
\includegraphics[clip,trim=0cm 0cm 0cm 0cm,width=\linewidth]{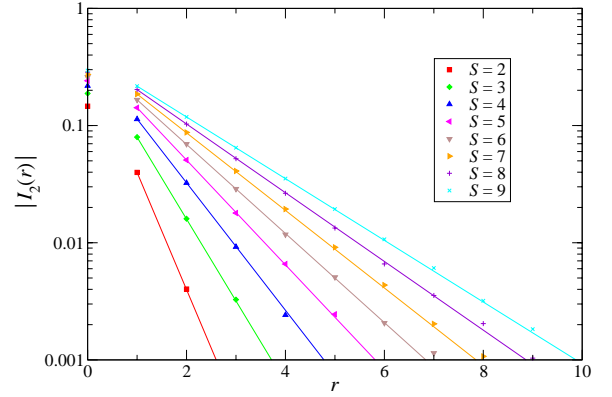}
\end{center}
\caption{\label{FIG:fig-chain-icorr} 
Reduced biquadratic spin correlations for spin-$S$ AKLT chains. Lines are exact forms taken from Ref.~\citenum{Freitag-ZPB-91} and given by Eq.~\eqref{EQ:two-spin-moments}, while data points are sampled according to the Monte Carlo scheme described in the main text.}
\end{figure}

We can also measure higher-order moments of the two-spin correlation function,
\begin{equation}
C_{n}(r) = \expectation{\bigl(\vec{S}_{0}\cdot\vec{S}_{r}\bigr)^{n}}.
\end{equation}
We introduce the notation $C_{n}$ and $\bar{C}_{n}$ to represent the normal and strange forms of the $n$th order moment of the correlation function obtained using pure and mixed expectation values, respectively. Using a transfer matrix method, Freitag and M\"{u}ller-Hartmann~\cite{Freitag-ZPB-91} were able to construct polynomials $I_{n}(r)$ for $1\le n\le S$ such that $I_{n}(r)$ depends on all $C_{m}(r)$ with $m\le n$, and with the following remarkable property (for $r\neq0$):
\begin{equation}\label{EQ:two-spin-moments}
I_{n}(r) = \frac{2n+1}{4\pi}(-1)^{nr}\Biggl(\frac{S!(S+1)!}{(S-n)!(S+n+1)!}\Biggr)^r.
\end{equation}
In Fig.~\ref{FIG:fig-chain-icorr} we compare the above analytic expression for $n=2$ to results from our Monte Carlo sampling scheme, where we make use of the relation $4\pi(S+1)^2(S+3)^2I_{2}(r)=30C_{2}(r)+15C_{1}(r)-10S^2(S+1)^2$ taken from the recursion relations given in Ref.~\citenum{Freitag-ZPB-91}.

\begin{figure}
\begin{center}
\includegraphics[clip,trim=0cm 0cm 0cm 0cm,width=\linewidth]{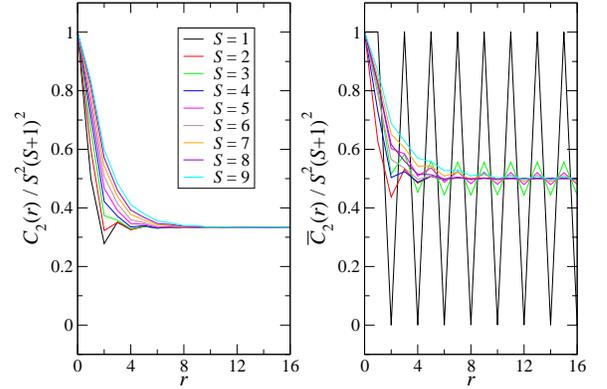}
\end{center}
\caption{\label{FIG:fig-chain-2corr} 
Normal (left) and strange (right) biquadratic spin correlations for spin-$S$ AKLT chains. For odd $S$, the strange biquadratic correlations exhibit oscillations with no decay in the oscillatory envelope.}
\end{figure}

In Fig.~\ref{FIG:fig-chain-2corr} we show biquadratic ($n=2$) spin correlations. For the normal estimator, we observe exponential decay to a constant value of $S^2(S+1)^2/3$, while for the strange estimator we observe exponential decay to a constant value of $S^2(S+1)^2/2$, but also see anomalous oscillations about this value (with no sign of decay in the envelope) for odd values of $S$.

We can also consider four-spin correlations of the form
\begin{equation}
F(r) = \expectation{\bigl(\vec{S}_{0}\cdot\vec{S}_{1}\bigr)\bigl(\vec{S}_{r}\cdot\vec{S}_{r+1}\bigr)},
\end{equation}
where as before we will use $F(r)$ and $\bar{F}(r)$ to refer to normal and strange correlations, respectively. However, we find both $F(r)$ and $\bar{F}(r)$ to quickly decay to the value $\expectation{\left(\vec{S}_{0}\cdot\vec{S}_{1}\right)}^{2}$ for all $S$, reflecting the fact that AKLT states are by definition fixed point valence bond solids with $M$ singlets per bond.

\subsubsection{Winding numbers}

Bonesteel has shown that short-ranged valence bond configurations can be classified by a topological invariant that counts the number of valence bonds cut by a vertical bond~\cite{Bonesteel-PRB-89}. For odd-width strips, vertical lines must cut through an even or odd number of valence bonds, and shifting the vertical line horizontally by one lattice spacing changes this number by $\pm1$. Thus, short-ranged valence bond states can be divided into two topologically distinct classes: one with an even-odd pattern and another with an odd-even pattern. Similarly, on even width strips the states break down into the sectors of all even or all odd cuts.

\begin{figure}
\begin{center}
\includegraphics{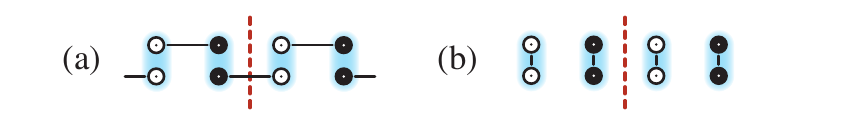}
\end{center}
\caption{\label{FIG:chain-sectors}
Topological sectors of short-ranged valence bond states on even-width strips: (a)~$S=1$ AKLT state in the odd sector and (b)~$S=1$ internal triplet state in the even sector.}
\end{figure}

The spin-$S$ AKLT chains considered in this paper can be thought of as short-ranged valence bond states in a strip geometry, with each position along the $x$ axis representing a single spin-$S$ site, and the constituent spin-1/2 flavors distributed along the $y$ axis. Since $2S$ is even for integral $S$, this corresponds to the even-width geometry discussed above, and valence bond states can be split into even and odd sectors. As illustrated in Fig.~\ref{FIG:chain-sectors}, the spin-$S$ AKLT states on a chain with odd $S$ all fall into the odd sector, while even $S$ AKLT states and internal triplet states belong to the even sector.

It can be shown that the valence bond overlap graphs between states in different topological sectors must have a nonzero winding number~\cite{Bonesteel-PRB-89}. Additionally, since the winding number may only change by $\pm2$, a topological winding number can be defined as the winding modulo 2. At large $S$, we have observed a small contribution to the strange winding for even $S$, which we interpret as a finite-size effect that should disappear in the thermodynamic limit. More to the point, we find that the topological winding is exactly zero for even $S$ AKLT states and exactly unity for odd $S$ AKLT states. Thus, the strange winding number modulo 2 is a topological invariant for AKLT states in one dimension.

As pointed out by Kim {\it et al.}, the winding number characterization of short-ranged valence bond configurations is intimately related to string order in one dimension~\cite{Kim-PRB-00}. The presence of string order implies a nonzero winding number, and vice versa. In fact, for any short-ranged valence bond configuration in one dimension, a string order parameter can be constructed in such a way as to be nonzero. Similarly, nonzero winding can be achieved with the choice of an appropriate reference state: for states in the odd sector an even reference state will yield nonzero winding, while for states in the even sector an odd reference state yields nonzero winding. Nonzero winding is also what gives rise to long-range strange correlations, since the estimator $\bra{\Omega}\vec{S}_{i}\cdot\vec{S}_{j}\ket{\psi}$ can only have nonzero values when $i$ and $j$ belong to the same loop in the transition graph overlap. Thus, we see that the topological winding numbers fully characterize the topological sectors of the AKLT states in one dimension, and also give rise to the strange correlator and the string order parameter.

\subsection{ \label{SSEC:2D} Honeycombs and squares}

There is an exact mapping of multiplicity $M$ AKLT states onto classical O(3) models at temperature $1/M$ {\it in the same number of dimensions}. In two dimensions, this mapping implies (via the Mermin-Wagner theorem) that AKLT states remain disordered for finite $M$. However, the correlation length is expected to grow exponentially with $M$. Additionally, an Orstein-Zernicke form is expected for the spin-spin correlation function, i.e., $C(r)\sim e^{-r/\xi}/r^{\eta}$. As we shall see, this makes it difficult to distinguish between algebraic and exponential asymptotic forms of the correlation function when $L<\xi$. As an alternative, we consider the winding fractions defined in Eq.~\eqref{EQ:winding-fraction}, which decay to zero as $L\rightarrow\infty$ whenever $C(r)$ decays exponentially and to a nonzero value whenever $C(r)$ decays algebraically.

In two dimensions, the boundary modes of SPT states are 1+1 dimensional gapless critical theories. Thus, they can be expected to be conformally invariant, and by extension the strange correlator should also be conformally invariant. Using this as our justification, we analyze the strange correlator using an algebraic function of the conformal length $\ell$; namely, $\bar{C}(\ell)\sim\ell^{-\eta}$, where $\ell=(L/\pi)\sin(\pi r/L)$.

\subsubsection{ \label{SSSEC:Honeycombs} Honeycombs}

The $S=3/2$ AKLT state on the honeycomb lattice is the first extension of AKLT states to dimension higher than 1. Early on, Affleck {\it et al.}~\cite{Affleck-CMP-88} and Kennedy {\it et al.}~\cite{Kennedy-JSP-88} were able to establish upper bounds on the correlation length ($\xi\leq3.54$ and $\xi\leq2.5$, respectively), which strongly implied a gapped disordered state. Later, Lou {\it et al.} calculated the entanglement spectrum, concluding that the boundary modes are related to a spin-1/2 Heisenberg ferromagnet with gapless quadratic dispersion~\cite{Lou-PRB-11}. More recently, Huang {\it et al.} have studied the $S=3/2$ AKLT state on the honeycomb lattice using the symmetry protected quantum renormalization group~\cite{Huang-PRB-13}. The $S=3/2$ AKLT state is a so-called weak SPT state that cannot be protected by on-site symmetry alone. Rather, it is protected by translational symmetry.

\begin{figure}
\begin{center}
\includegraphics[clip,trim=0cm 0cm 0cm 0cm,width=\linewidth]{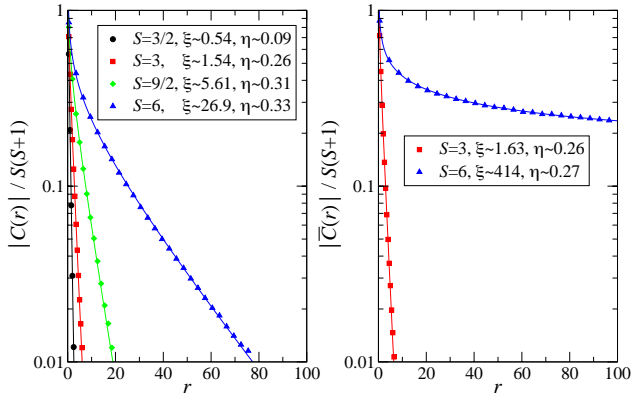}
\end{center}
\caption{\label{FIG:fig-honeycomb}
Normal (left) and strange (right) spin correlation functions for spin-$S$ AKLT states on a honeycomb lattice with dimensions $384\times256$. The strange correlator is calculated using the mixed overlap with an internal triplet product state, which is only defined for integer values of $S$. Lines are fits to a functional form $C(r)=f(r)+f(L-r)$, with $f(r)\sim\exp(-r/\xi)/r^{\eta}$. Correlations are shown along the zigzag direction and are nearly isotropic. All distances are given in units of the honeycomb primitive cell.}
\end{figure}

In Fig.~\ref{FIG:fig-honeycomb} we show the normal and strange correlator for spin-$S$ AKLT states on the honeycomb lattice. The normal correlator decays exponentially with a very small correlation length, even for $S=3$ (the correlation length is expected to grow as $\xi\sim e^{\pi M}$ in 2D spin-$S$ AKLT states~\cite{Arovas-PRL-88}, with $M=2S/z$ the multiplicity and $z=3$ the lattice coordination). The strange correlator is calculated using the internal triplet product state, which can only be defined for $S$ an integer (in general, no on-site symmetric product state can be formed for fractional $S$). For $S=3$, we see that the strange correlator also displays exponential decay with a correlation length similar to that of the normal correlator. Because the $S=3$ AKLT state can be thought of as two coupled copies of the $S=3/2$ AKLT state, we expect it to be a trivial state (similar to the even-odd effect in spin-$S$ AKLT chains). We are not able to see the expected exponential decay for the $S=6$ strange correlator, which we believe is due to finite-size effects (in other words, the correlation length is larger than the maximum system size studied, $L=1024$). This is complicated by the Ornstein-Zernike form of the correlations, $\exp(-r/\xi)/r^{\eta}$, which for $r\ll\xi$ is indistinguishable from power-law decay.

\begin{figure}
\begin{center}
\includegraphics[clip,trim=0cm 0cm 0cm 0cm,width=\linewidth]{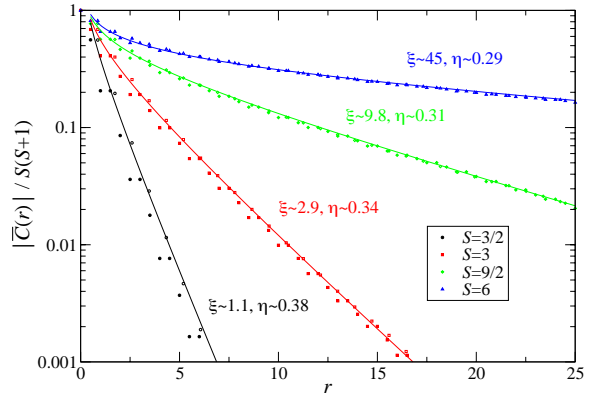}
\end{center}
\caption{\label{FIG:fig-honeycomb-kekule}
Strange correlator for spin-$S$ AKLT states on a honeycomb lattice with dimensions $384\times256$, as obtained using a pure dimer product state in the kekule pattern. Solid symbols denote distances along the zigzag direction, while empty symbols stand for the armchair direction. Lines are fits to a functional form $C(r)=f(r)+f(L-r)$, with $f(r)\sim\exp(-r/\xi)/r^{\eta}$. All distances are given in units of the honeycomb primitive cell.}
\end{figure}

Although we cannot form an on-site symmetric product state for half-odd-integer $S$, there are many possible dimer product states that retain the full spin rotational symmetry. However, a complication arises if we wish to use a dimer product state as the reference state for calculating the strange correlator. This is because dimer product states on a bipartite lattice can be classified by {\it dimer winding numbers} that lead to nonzero winding (and thus long-range correlations) between states in different {\it dimer winding sectors}~\cite{Albuquerque-PRB-10,Patil-PRB-14}. Thus, in order to obtain an unambiguous signal from the strange correlator, we must choose a dimer product state that is in the same dimer winding sector as the AKLT state we wish to probe. Following Ref.~\citenum{Patil-PRB-14}, we define dimer winding numbers along the three zigzag axes of the honeycomb lattice $w_{x}$, $w_{y}$, and $w_{z}$ as the number of valence bonds that are crossed in these three directions. It turns out the only dimer pattern that fits this description is the kekule pattern, but as shown in Fig.~\ref{FIG:fig-honeycomb-kekule}, the strange correlator is short-ranged using such a reference state. This is due to the fact that the kekule dimerization breaks translational symmetry, which is a protecting symmetry of the AKLT states in two dimensions (i.e., they are ``weak'' SPTs).

\begin{figure}
\begin{center}
\includegraphics[clip,trim=0cm 0cm 0cm 0cm,width=\linewidth]{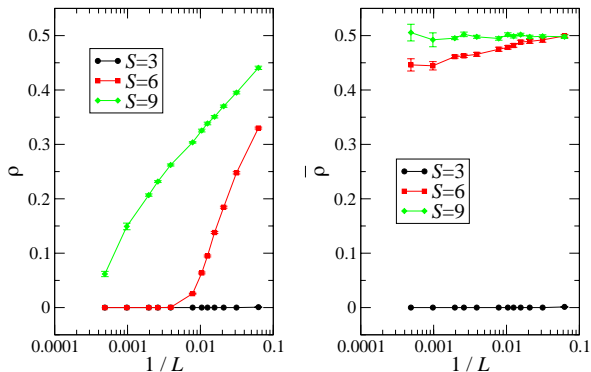}
\end{center}
\caption{\label{FIG:fig-wind-honeycomb}
Normal (left) and strange (right) winding fractions of spin-$S$ AKLT states on the honeycomb lattice.}
\end{figure}

As in one dimension, winding number fractions are an alternative way to search for (quasi-)long-range correlations in two dimensions. In Fig.~\ref{FIG:fig-wind-honeycomb}, we show these quantities for even multiplicity AKLT states on the honeycomb lattice. As expected, $\rho$ approaches zero with increasing system size. For small $S$, $\bar{\rho}$ either approaches zero ($S=3$), or shows signs of decreasing at larger system sizes ($S=6$). This allows us to confirm that even multiplicity states are trivial for $S=3$ and $S=6$. For $S=9$, we also expect $\bar{\rho}$ to ultimately decay to zero, but at system sizes beyond our present computational capabilities (our largest system has a total of $N\times2S=2\times2048^{2}\times18=150\, 994\, 944$ flavor-site combinations).

\subsubsection{ \label{SSSEC:Squares} Squares }

We now move on to the AKLT state on a square lattice, which has been considered by several past studies. Early on, Kennedy {\it et al.} were able to provide a rigorous upper bound to the correlation length $\xi<160$~\cite{Kennedy-JSP-88}. Later, Lou {\it et al.} calculated the entanglement spectrum, concluding that the boundary modes are related to a spin-1/2 Heisenberg antiferromagnet with gapless linear dispersion and central charge $c=1$~\cite{Lou-PRB-11}. You {\it et al.} calculated the strange correlator for this state using a combination of transfer matrix and DMRG, finding an algebraic decay with exponent $\eta\simeq0.32$~\cite{You-PRL-14}. The spin-2 AKLT state on the square lattice has also been proposed as a universal resource for measurement-based quantum computation~\cite{Wei-PRA-15}. Finally, a recent proposal for classifying SPT phases with mirror reflection and on-site ${\mathbb Z}_{2}$ symmetry includes the AKLT states within its classification scheme~\cite{Yoshida-PRB-15}.

\begin{figure}
\begin{center}
\includegraphics[clip,trim=0cm 0cm 0cm 0cm,width=\linewidth]{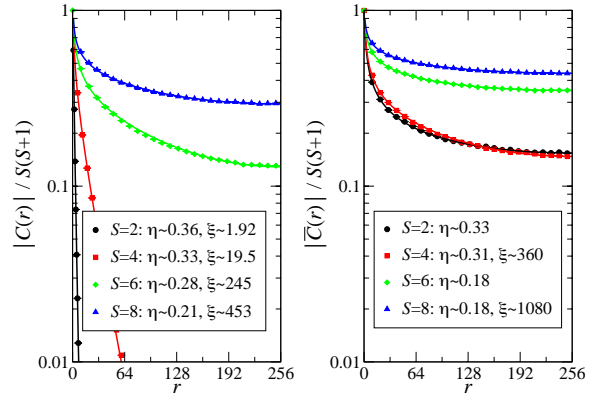}
\end{center}
\caption{\label{FIG:fig-square}
Normal and strange correlators for the spin-$S$ AKLT state on the square lattice with $S\le8$. Both correlators are very nearly isotropic (data shown along the $x$ direction). Solid lines for the normal correlator and the strange correlator with even $M$ are from fits to the form $C(r)=f(r)+f(L-r)$ with $f(r)\sim\exp(-r/\xi)/r^{\eta}$, taking into account the periodic boundaries of the $L=512$ simulation cell. For the odd $M$ strange correlator, fits are obtained using an algebraic function of the conformal length, $C(\ell)\sim\ell^{-\eta}$.}
\end{figure}

In Fig.~\ref{FIG:fig-square} we show the normal and strange correlations for spin-$S$ AKLT states on the square lattice. By fitting the correlations to an exponential form that takes into account the periodicity of the lattice, $C(r)=f(r)+f(L-r)$, with $f(r)\sim\exp(-r/\xi)/r^{\eta}$, we see that normal correlations decay exponentially, but with a correlation length $\xi$ that quickly grows with the spin quantum number [we note that $\xi\sim\exp(\pi M)$ is expected from theoretical considerations~\cite{Arovas-PRL-88}, with $M=2S/z$ the multiplicity of the AKLT state with lattice coordination $z$]. For the strange correlator with even $M$, an exponential fit also works, but the correlation lengths (with the exception of the $S=4$ data) exceed the size of our simulation cell ($L=512$). Thus, we cannot distinguish between exponential and algebraic decay without resorting to larger system sizes. For odd $M$, we fit the strange correlator using an algebraic function of the conformal length, $\bar{C}(\ell)\sim\ell^{-\eta}$. Importantly, this functional form does not work well for the $S=4$ (even $M$) state. This is related to the fact that the strange correlator decreases more rapidly for the $S=4$ AKLT state than for the $S=2$ AKLT state, which is an indication that the $S=4$ AKLT state has a finite correlation length.

\begin{figure}
\begin{center}
\includegraphics[clip,trim=0cm 0cm 0cm 0cm,width=\linewidth]{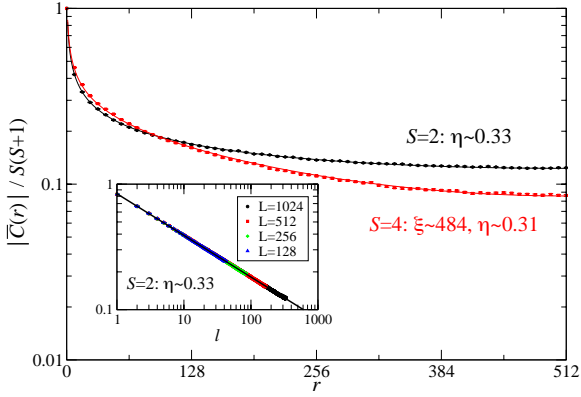}
\end{center}
\caption{\label{FIG:fig-square-evenodd}
Strange correlators for the $S=2$ and $S=4$ AKLT states on a square lattice with $L=1024$. Both correlators are very nearly isotropic, and data are shown along the $x$ direction. Solid lines for the $S=2$ and $S=4$ data are fits to the conformal and periodic forms mentioned in the main text. Inset: data for $S=2$ shown on a log-log scale demonstrates size-independent algebraic decay when plotted as a function of the conformal length $\ell$.}
\end{figure}

To investigate the difference between the $S=2$ and $S=4$ AKLT states on the square lattice in greater detail, we look at the strange correlator for system sizes $L=1024$ (with $N=L^{2}$ nearly one million, close to the maximum system size we can simulate). As shown in Fig.~\ref{FIG:fig-square-evenodd}, fitting to the $S=2$ data yields a decay exponent that is identical to the $L=512$ result. However, the $S=4$ data is now best fit by a correlation length $\xi\sim 468<L$, indicating a short-ranged strange correlator and trivial SPT character. This in turn implies an even-odd effect for AKLT states with multiplicity $M=2S/z$ (for $S=2$ the multiplicity is 1 while $S=4$ has multiplicity 2), in agreement with a recent classification scheme for bosonic SPT states in two dimensions protected by reflection symmetry~\cite{Yoshida-PRB-15}.

In principle, the power-law exponent of the strange correlator should match that of the boundary theory. For the case of the square lattice AKLT state with odd multiplicity, the edge theory should be given by a conformal field theory with central charge $c=1$ and Luttinger parameter $K=1$. In our present work, we have found $\eta\approx0.33$, which does not match the expected $1/r$ decay predicted from the Luttinger theory of a half-odd-integral Heisenberg spin chain. Note, however, that the internal triplet state reduces the O(3) spin symmetry down to $\text{O}(2)\times {\mathbb Z}_{2}$. Thus, we should not expect a direct relation to the Luttinger theory of the isotropic Heisenberg chain unless we instead choose an O(3) symmetric product state.

\begin{figure}
\begin{center}
\includegraphics[clip,trim=0cm 0cm 0cm 0cm,width=\linewidth]{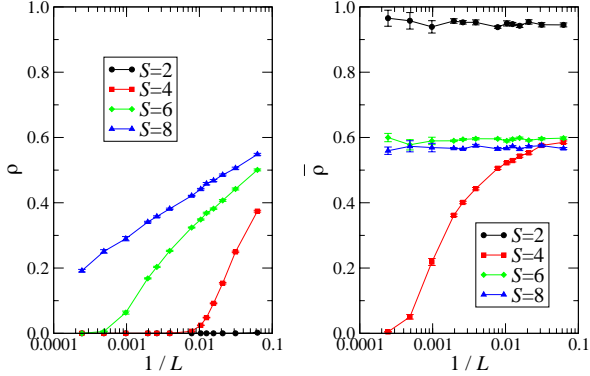}
\end{center}
\caption{\label{FIG:fig-wind-square}
Normal (left) and strange (right) winding fractions of spin-$S$ AKLT states on the square lattice.}
\end{figure}

As before, we can also define winding numbers for the AKLT on the square lattice. In contrast to the AKLT chain, the topological component of the winding numbers for mixed overlaps is now strictly zero---that is, the square lattice AKLT state and the internal triplet state are in the same topological sector as defined by the topological winding. However, the nontopological component of the winding numbers is nonzero for the mixed overlap state. In particular, we find a finite strange winding fraction $\bar{\rho}\simeq0.95$, whereas the normal winding fraction $\rho=0$ in keeping with the lack of N\'{e}el order and hence zero spin stiffness in the square lattice AKLT state. These results are summarized in Fig.~\ref{FIG:fig-wind-square}, where we also show data for larger values of $S$. The even-odd effect is clearly seen in the downward turn of the strange winding fraction for $S=4$, whereas for $S=8$ we are not able to reach large enough system sizes to see this trend.

\subsection{ \label{SSEC:3D} Cubes, Diamonds, and Hyperhoneycombs }

\begin{figure}
\begin{center}
\includegraphics[clip,trim=0cm 0cm 0cm 0cm,width=\linewidth]{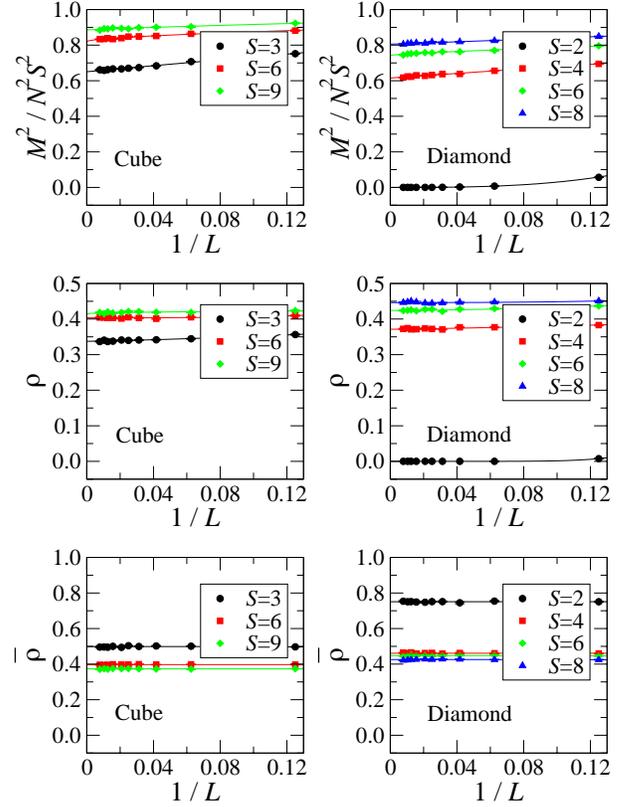}
\end{center}
\caption{\label{FIG:fig-wind-cube-diamond}
Normal squared magnetization $M^{2}$ (top), normal winding fraction $\rho$ (middle) and strange winding fraction $\bar{\rho}$ (bottom) of spin-$S$ AKLT states on the cubic (left) and diamond (right) lattices. In all panels, values are extrapolated to the $L\rightarrow\infty$ limit by fits to the form $a+b/L^{c}$ (solid lines).}
\end{figure}

To investigate AKLT states on 3D lattices, we use the winding fractions as defined in Eq.~\eqref{EQ:winding-fraction}. Results for the cubic and diamond lattices are shown in Fig.~\ref{FIG:fig-wind-cube-diamond}. On the cubic lattice, $\rho$ indicates that magnetic order is present beginning with the multiplicity one AKLT state at $S=3$. In contrast, $\rho$ is zero for the multiplicity one $S=2$ AKLT state on the diamond lattice, with magnetic order setting in for $S\geq4$ (or, $M\geq2$). These results are in agreement with Monte Carlo investigations of the equivalent classical model at inverse temperature $M$, which found magnetic order on the cubic and diamond lattices for $M\gtrsim0.60$ and $M\gtrsim1.18$, respectively~\cite{Parameswaran-PRB-09}. Our present study goes one step further to show that, in addition to being magnetically disordered, the $M=1$ AKLT state on the diamond lattice is also a nontrivial SPT state.

\begin{figure}
\begin{center}
\includegraphics[clip,trim=0cm 0cm 0cm 0cm,width=\linewidth]{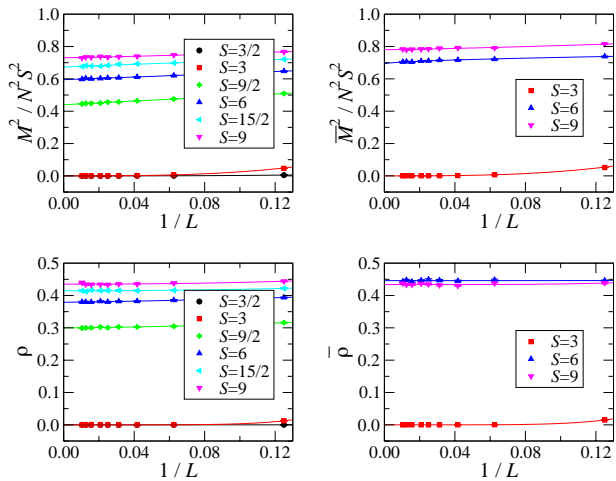}
\end{center}
\caption{\label{FIG:fig-wind-hyperhoneycomb}
Normal (left) and strange (right) values for the squared magnetization (top) and winding fractions (bottom) of spin-$S$ AKLT states on the hyperhoneycomb lattice. In all panels, values are extrapolated to the $L\rightarrow\infty$ limit by fits to the form $a+b/L^{c}$ (solid lines).}
\end{figure}

We also consider the hyperhoneycomb lattice, a tricoordinated lattice with elementary loops of ten lattice sites and four sites per primitive cell. We are not aware of any study of AKLT states on this lattice, but due to its lower connectivity (as compared with the cubic and diamond lattices) we can expect that at least the $M=1$ state is disordered. As shown in Fig.~\ref{FIG:fig-wind-hyperhoneycomb}, we actually find zero winding in the normal overlap for both the $M=1$ and $M=2$ states, which indicates a nonmagnetic (quantum paramagnetic) ground state. For $M\ge3$, the winding fraction is stable with respect to system size as the ground state develops long-range magnetic order. For even multiplicity, we can also measure winding in the mixed overlap between the AKLT state and the internal triplet product state. For $M=2$ we find zero winding, indicating a trivial quantum paramagnet and confirming the even-odd effect in 3D AKLT states, while for $M\ge4$ we find a nonzero winding fraction, which in this case reflects the broken symmetry of the magnetically ordered ground state.

\section{\label{SEC:energy} Energy Gaps}

In one dimension, AKLT states possess a finite gap to excitations~\cite{Affleck-CMP-88}. In two dimensions, AKLT states are also believed to be gapped; this has not be proved rigorously, although for the honeycomb and square lattices AKLT states, all correlation functions have been shown to decay exponentially~\cite{Kennedy-JSP-88}.

Much progress has been made on the numerical front. Early on, Arovas {\it et al.} used the single-mode approximation to obtain an upper bound of the excitation gap $\Delta\le\frac{10}{27}\simeq0.370$ for the $S=1$ AKLT chain~\cite{Arovas-PRL-88}. More recently, Garcia-Saez {\it et al.} used tensor network methods to obtain an estimate $\Delta\approx0.350$ for the $S=1$ AKLT chain, as well as values of $\Delta\approx0.10$ and $\Delta\approx0.015$ (in our units) for the $S=3/2$ AKLT honeycomb and $S=2$ AKLT square lattices, respectively~\cite{Garcia-Saez-PRB-13}.

Ganesh {\it et al.} have also estimated the excitation gap for the $S=3/2$ AKLT honeycomb using exact diagonalization, finding a value $\approx0.1$ consistent with the literature~\cite{Ganesh-PRB-11}, while Poilblanc {\it et al.} have found a critical field $\simeq 0.113$ for the $S=3/2$ AKLT honeycomb, in agreement with the expectation that the field must close the gap~\cite{Poilblanc-PRB-13}. More recently, Vanderstraeten {\it et al.} have used a variational ansatz based on the PEPS formalism to find an upper bound $\Delta_{\text{SMA}}=0.0199$ and variational estimate $\Delta_{\text{var}}=0.0147$ for the excitation gap of the $S=2$ AKLT state on the square lattice, as well as a rather precise value for the correlation length $\xi=2.06491$~\cite{Vanderstraeten-PRB-15}.

Excitation gaps can also be estimated using the valence bond formalism discussed in this paper. In this formalism, the most natural excitation is to convert a singlet bond into a triplet. In particular, a superposition of AKLT states with the triplet in all possible locations, $\ket{\psi_t}=\sum_{\langle i,j \rangle}(S_i^z-S_j^z)\ket{\psi}$, is a very good approximation for obtaining the smallest energy gap [at $\pi$ for a chain, $(\pi,\pi)$ for a square, etc.].

In Table~\ref{TAB:gaps} we show energy gap estimates obtained from $\Delta_\text{est}=\bra{R}\hat{H}\ket{\psi_t}/\overlap{R}{\psi_t}$, where in this case the reference state $\ket{R}$ is chosen to be the N\'{e}el state that has equal overlap with all valence bond configurations. Note that since $\Delta_\text{est}$ is  obtained from a projection instead of an expectation value, it is not a variational estimate.

We also attempted to access the excitation gap asymptotically using the projected state ${\hat H}^{m}\ket{\psi_{t}}$. However, due to the presence of a quantum Monte Carlo sign problem we were unable to obtain sufficiently converged results to present here. Since the sign problem for AKLT Hamiltonians is not expected to be severe (there are no explicitly frustrated interactions), the failure of our projected triplet states to converge is probably due to limitations in our sampling scheme. To overcome these limitations, it would be interesting to develop a generalization of the loop algorithm in Ref.~\citenum{Sandvik-PRB-10} to the case of spin-$S$ representations of valence bond states. However, such a study is outside the scope of the present paper and we leave it for future consideration.

\begin{table*}
\caption{Energy gaps for various lattices. $\Delta_\text{est}$ are the energy gap estimates obtained in this work, while $\Delta$ are energy gap measurements obtained from the indicated references. The Hamiltonian parameters in Eq.~\eqref{EQ:AKLT_Hamiltonian} are chosen as $A_{J}=1$.}
\begin{tabular}{  l  l  l  l  S[table-format=3.7]  S[table-format=3.4]  }
\hline \hline
lattice & \!\!\!\!\!\!dimension~~~~~~ & \!\!\!\!\!\!multiplicity~~ & ~spin & $\Delta_\text{est}$ & $\Delta$ \\[+0.05cm]
\hline
chain & $d=1$ & $M=1$ & $S=1$ & 0.333333 & 0.350\!\!\textsuperscript{a} \\[-0.15cm]
chain & $d=1$ & $M=2$ & $S=2$ & 0.1 & \\[+0.1cm]
honeycomb & $d=2$ & $M=1$ & $S=3/2$ & 0.1 & 0.113\!\!\textsuperscript{b} \\[-0.15cm]
square & $d=2$ & $M=1$ & $S=2$ & 0.0285714 & 0.0147\,\textsuperscript{c} \\[+0.1cm]
hyperhoneycomb~~~ & $d=3$ & $M=1$ & $S=3/2$ & 0.1 & \\[-0.15cm]
diamond & $d=3$ & $M=1$ & $S=2$ & 0.0285714 & \\[+0.05cm]
\hline \hline
{}$^\text{a}$~from Ref.~\citenum{Garcia-Saez-PRB-13}\\[-0.15cm]
{}$^\text{b}$~from Ref.~\citenum{Poilblanc-PRB-13}\\[-0.15cm]
{}$^\text{c}$~from Ref.~\citenum{Vanderstraeten-PRB-15}
\end{tabular}
\label{TAB:gaps}
\end{table*}

\section{\label{SEC:Conclusions} Conclusions }

We have developed a numerical framework for computing the strange correlator in the valence bond basis, which allows us to characterize the symmetry protected topological properties of AKLT states in one, two and three dimensions. In one dimension, our method gives a clear picture of the origin of long-range correlations in the strange correlator (viz., the persistence of a single system-spanning valence bond loop). It also allows us to make connections between the strange correlator and the topological winding number sectors~\cite{Bonesteel-PRB-89}. In two and three dimensions, the topological winding number characterization disappears, but we were able to show that a winding number fraction (in analogy to the superfluid fraction or spin stiffness) survives and can be used as an alternative way to characterize SPT states.

In one dimension, comparison to exact results confirms the accuracy of our numerical method. Notably, our results reproduce the well-known even-odd effect in 1D AKLT chains: states with even integer spin ($S$ even) are trivial states, whereas  odd integer spin states ($S$ odd) are nontrivial SPT states. In two dimensions, we demonstrate that the same even-odd effect exists on the honeycomb and square lattices, where even and odd now refer to the multiplicity $M$ of the AKLT state ($M$ even is trivial, $M$ odd is nontrivial). In three dimensions, we observe the signatures of magnetic order for large spin, confirming past Monte Carlo studies of the equivalent classical O(3) model~\cite{Parameswaran-PRB-09}. We were also able to provide direct evidence of nontrivial SPT character for the $S=2$ diamond AKLT state and to show that the $S=3/2$ and $S=3$ hyperhoneycomb AKLT states do not magnetically order. 

We have also applied the valence bond formalism to calculate energy gap estimates. By sequentially operating the full Hamiltonian on a trial state in the one-triplet sector, we were able to estimate the triplet gap. Our values compare well with results from other methods. In principle a sign-sampled quantum Monte Carlo method should be able to asymptotically approach the exact energy gap; however, we were unable to obtain sufficiently converged results using a naive sampling method for the Hamiltonian projection ${\hat H}^{m}\ket{\psi_{t}}$.

The methods employed in this paper can also be applied to a wide variety of SPT states away from the exact AKLT points. This includes weakly coupled Haldane chains~\cite{Wierschem-MPL-14}, which have drawn renewed interest in light of the new SPT classification scheme, as well as ``disordered'' versions of the AKLT states where defects such as domain walls are allowed to proliferate, which may be a way to study the topological phase transition from weak to strong SPT phases~\cite{You-arXiv-16}.

\acknowledgments

The authors acknowledge support from the Office of Research and Sponsored 
Programs of the University of Mississippi. One of the authors (K.S.D.B.) benefited
from a stay at the Aspen Center for Physics under NSF Grant No.\ PHY-1066293.

\appendix

\section{\label{SEC:AppA}Rules for loop updates}

During the Monte Carlo sampling of valence bond overlaps, it is necessary to evaluate the relative weights of various loop structures. Fortunately, these are easily calculated in the valence bond basis and are determined solely by $N_{\ell}$, the number of loops formed by a particular valence bond overlap. Thus, calculation of the relative weight of valence bond overlaps amounts to counting the relative difference in $N_{\ell}$ before and after a proposed update, i.e. $\Delta N_{\ell}=N_{\ell}^\text{new}-N_{\ell}^\text{old}$.

In this appendix, we construct a set of rules for counting $\Delta N_{\ell}$ for a generic bipartite loop under two types of updates: on-site symmetrization of the flavor components and operation of the singlet projector operator.

We introduce a singlet projection operator $\hat{P}_{\mu,\nu} = \frac{1}{4} - \vec{s}_{\mu} \cdot \vec{s}_{\nu}$ and a parity operator as $\hat{Q}_{\mu,\nu} = \frac{1}{4} + \vec{s}_{\mu} \cdot \vec{s}_{\nu}$, where $\mu$ and $\nu$ are spin-$\frac{1}{2}$ flavors residing at sites $i_{\mu}$ and $j_{\nu}$, respectively. These will allow us to keep track of rearrangements of our bipartite valence bond states through the following useful formulas:
\begin{equation}
\begin{split}
\hat{P}_{\mu,\nu} \ket{[\mu,\nu]}       &=  +\ket{[\mu,\nu]} \\ 
\hat{P}_{\nu\sigma} \ket{[\mu,\nu][\sigma,\tau]}  &=  +\frac{1}{2} \ket{[\mu,\tau][\sigma,\nu]} \\ 
2\hat{Q}_{\mu,\nu} \ket{[\mu,\nu]}       &= -\ket{[\mu,\nu]} \\ 
 \hat{Q}_{\nu\tau} \ket{[\mu,\nu][\sigma,\tau]}  &=  +\frac{1}{2} \ket{[\mu,\tau][\sigma,\nu]} 
\end{split}
\end{equation}
where each singlet $\ket{[\mu,\nu]}\equiv\bigl(\ket{\uparrow}_{\mu}\otimes\ket{\downarrow}_{\nu} - \ket{\downarrow}_{\mu}\otimes\ket{\uparrow}_{\nu}\bigr)/\sqrt{2}$ is directed from sublattice A to sublattice B (that is, $i_{\mu}\in\text{A}$ and $j_{\nu}\in\text{B}$).

When calculating the strange correlator, we use an on-site product state consisting of the symmetric superposition of triplets, with $\ket{\{\mu,\nu\}}\equiv\bigl(\ket{\uparrow}_{\mu}\otimes\ket{\uparrow}_{\nu} + \ket{\downarrow}_{\mu}\otimes\ket{\downarrow}_{\nu}\bigr)/\sqrt{2}$. On bipartite lattices, the resulting loops in the transition graph must contain an even number of triplet links, which allows us to perform the usual operations without having to worry about potential singlet-triplet annihilation (in general, two valence bond states are orthogonal if any of the loops in their overlap graph contain an odd number of triplets). It is useful to list the modified formulas for $\hat{P}$ and $\hat{Q}$ in this sector:
\begin{equation}
\begin{split}
 \hat{P}_{\mu,\nu} \ket{\{\mu,\nu\}}     &=  \phantom{+}0 \\ 
 \hat{P}_{\nu,\sigma} \ket{\{\mu,\nu\}\{\sigma,\tau\}}  &=  -\frac{1}{2} \ket{[\mu,\tau][\sigma,\nu]} \\ 
2\hat{Q}_{\mu,\nu} \ket{\{\mu,\nu\}}     &= +\ket{\{\mu,\nu\}} \\ 
 \hat{Q}_{\nu,\tau} \ket{\{\mu,\nu\}\{\sigma,\tau\}} &=  +\frac{1}{2} \ket{\{\mu,\tau\}\{\sigma,\nu\}}. 
\end{split}
\end{equation}
Let us note that the minus sign in the above equation does not lead to an overall sign problem in the Monte Carlo simulation scheme employed in this paper. This is due to the fact that the overlap between a singlet valence bond configuration $\ket{v}$ and a triplet configuration $\ket{t}$ is always in the same sign sector, which is given by $(-1)^{SN/2}$. On a bipartite lattice, $N$ is always even, while the triplet state $\ket{t}$ is only defined for integer $S$.

\subsection{Singlet projection operator updates}

\begin{figure}
\begin{center}
\includegraphics[clip,trim=0cm 0cm 0cm 0cm,width=\linewidth]{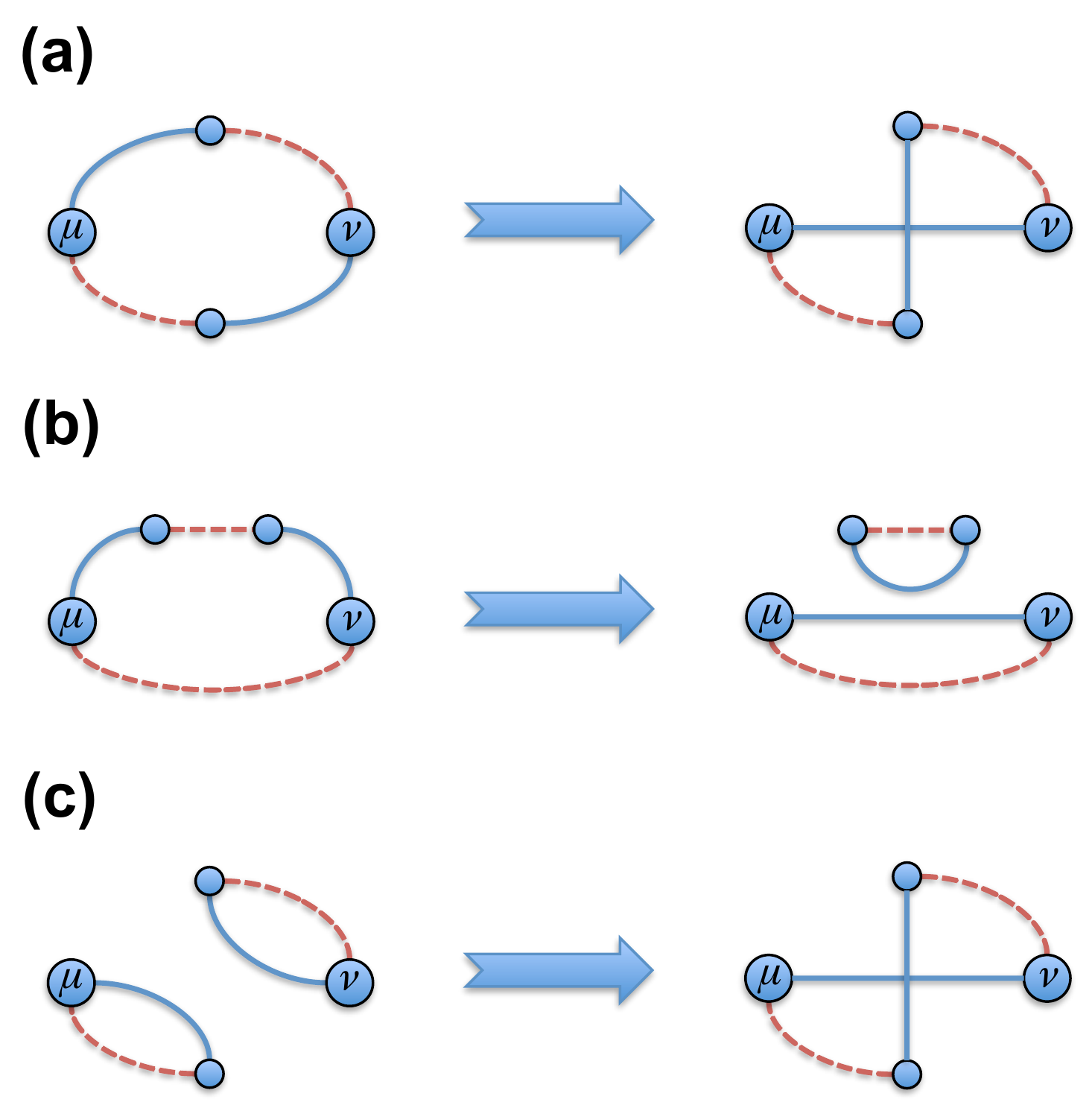}
\end{center}
\caption{\label{FIG:loop-rules-singlet}
Possible rearrangements for singlet updates. (a) $\mu$ and $\nu$ belong to the same loop with an even number of valence bonds connecting them. After the update, they remain in the same loop with no change in the loop ordering structure. (b) $\mu$ and $\nu$ belong to the same loop with an odd number of valence bonds connecting them. After the update, the loop is split so that $\mu$ and $\nu$ no longer belong to the same loop. (c) $\mu$ and $\nu$ belong to separate loops that merge into a single loop, after which $\mu$ and $\nu$ have an odd loop spacing.}
\end{figure}

In a singlet projector update, we wish to act upon a pair of flavor components with a singlet projector. In this case the pair of flavors $\mu$, $\nu$ belong to different sites. If $\mu$ and $\nu$ belong to the same sublattice, we instead use the parity operator $\hat{Q}$. When this occurs, one of three possible rearrangements will occur (see Fig.~\ref{FIG:loop-rules-singlet}).

First, if $\mu$ and $\nu$ belong to separate loops, these loops will merge, and $\Delta N_{\ell}=-1$. Next, if $\mu$ and $\nu$ belong to the same loop, this loop will either remain intact or split in two, depending on the ``loop distance'' (call this $\Delta\ell$) between $\mu$ and $\nu$. If $\Delta\ell$ is even, the loop remains intact and $\Delta N_{\ell}=0$. If $\Delta\ell$ is odd, the loop splits in two as long as $\mu$ and $\nu$ do not belong to the same singlet, and $\Delta N_{\ell}=1$. In the case where $\mu$ and $\nu$ already form a singlet with one another, the loop obviously remains intact, even though $\Delta\ell=1$ by definition.

Combining the prefactors of the singlet projection operator (1/2 or 1 depending on whether or not $\mu$ and $\nu$ belong to different singlets) with the above determined loop weights, we obtain the total weights for singlet projection operator updates as follows:
\begin{itemize}
\item[(i)] If $\mu$ and $\nu$ belong to different loops, then combine loops with probability 1/4.
\item[(ii)] If $\mu$ and $\nu$ belong to the same loop with even loop spacing, then keep the reordered loop with probability 1/2.
\item[(iii)] If $\mu$ and $\nu$ belong to the same loop with odd loop spacing and are not connected as a singlet, then split the loops with probability 1.
\item[(iv)] If $\mu$ and $\nu$ belong to the same loop and are connected as a singlet, then keep the current loop with probability 1.
\end{itemize}

\subsection{Parity operator updates}

\begin{figure}
\begin{center}
\includegraphics[clip,trim=0cm 0cm 0cm 0cm,width=\linewidth]{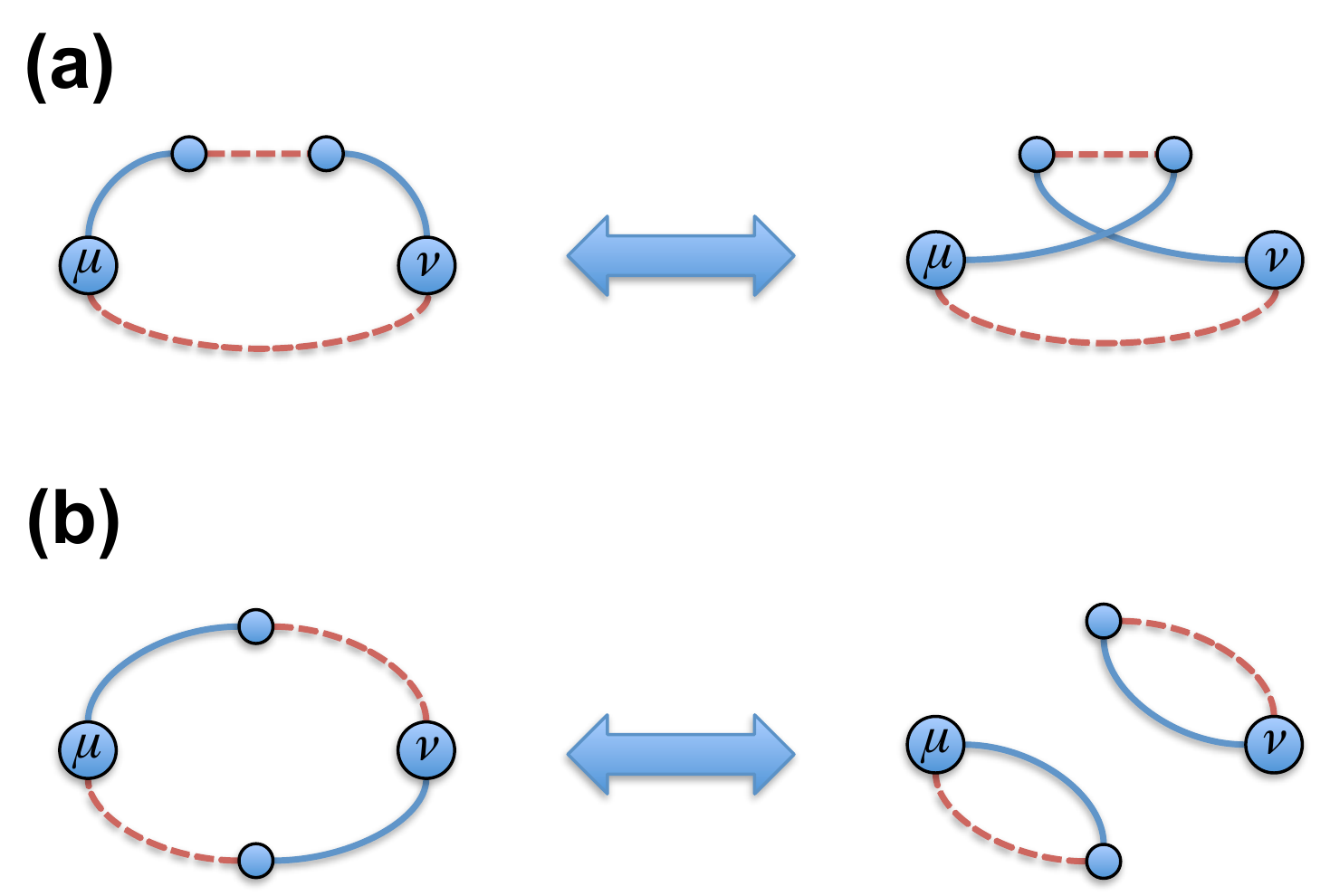}
\end{center}
\caption{\label{FIG:loop-rules-parity}
Possible rearrangements for parity updates. (a) $\mu$ and $\nu$ belong to the same loop with an odd number of valence bonds connecting them. After the update, they remain in the same loop while the two untouched segments (red dashed lines) retain their internal structure, but pick up an odd factor in their relative ordering. (b) $\mu$ and $\nu$ belong to the same loop with an even number of valence bonds connecting them. After the update, the loop is split so that $\mu$ and $\nu$ no longer belong to the same loop. In the reverse update, $\mu$ and $\nu$ belong to separate loops that merge into a single loop, after which $\mu$ and $\nu$ have an even loop spacing.}
\end{figure}

When the flavors $\mu$ and $\nu$ belong to the same sublattice, we instead use the parity operator $\hat{Q}$. In this case, one of three possible rearrangements will occur (see Fig.~\ref{FIG:loop-rules-parity}).

As before, if $\mu$ and $\nu$ belong to separate loops, these loops will merge, and $\Delta N_{\ell}=-1$. Next, if $\mu$ and $\nu$ belong to the same loop, this loop will either remain intact or split in two, depending on the ``loop distance'' (call this $\Delta\ell$) between $\mu$ and $\nu$. If $\Delta\ell$ is even, the loop splits in two, and $\Delta N_{\ell}=1$. If $\Delta\ell$ is odd, the loop remains intact and $\Delta N_{\ell}=0$. Note that the loop update rules are the opposite of the singlet projection operator for the case where $\mu$ and $\nu$ belong to the same loop.

Combining the prefactors of the singlet projection operator (1/2 or 1 depending on whether or not $\mu$ and $\nu$ belong to different singlets) with the above determined loop weights, we obtain the total weights for singlet projection operator updates as follows:
\begin{itemize}
\item[(i)] If $\mu$ and $\nu$ belong to different loops, then combine the loops with probability 1/4.
\item[(ii)] If $\mu$ and $\nu$ belong to the same loop with even loop spacing, then split the loops with probability 1.
\item[(iii)] If $\mu$ and $\nu$ belong to the same loop with odd loop spacing, then keep the reordered loop with probability 1/2.
\end{itemize}

\subsection{Symmetrization updates}

In a symmetrization update, we wish to symmetrize over all possible relabelings of the on-site flavor components. This entails randomly picking a pair of flavors on a given site (call them $\mu$ and $\nu$), then exchanging their labels. This update is related to the parity operator update, but in this case the factor of $\frac{1}{2}$ that comes from acting on a state with $\hat{Q}$ will drop out in the detailed balance equation; only the weight associated with loop rearrangements survives. (Equivalently, we can consider symmetrization updates as acting upon the current state with an operator $2\hat{Q}$.)
As discussed above, one of three possible rearrangements will occur. This time, their total probabilities are as follows:
\begin{itemize}
\item[(i)] If $\mu$ and $\nu$ belong to different loops, then combine the loops with probability 1/2.
\item[(ii)] If $\mu$ and $\nu$ belong to the same loop with even loop spacing, then split the loops with probability 1 (relative weight is 2).
\item[(iii)] If $\mu$ and $\nu$ belong to the same loop with odd loop spacing, then keep the reordered loop with probability 1.
\end{itemize}

Thus, to construct the symmetrization updates appears to be quite simple. Just choose some $\mu$ and $\nu$, then use $2\hat{Q}_{\mu\nu}$ to reorder the valence bond pattern with probability $p$ given by the above list. Since $p$ depends on whether or not $\mu$ and $\nu$ belong to the same loop, at first glance it appears that we have to keep a record of all loops. Actually, this is not the case, and we can in fact form a ``fast'' update that does not need to keep track of loop memberships during the course of the update. The trick is to take advantage of the bipartite nature of valence bonds and construct a pseudospin degree of freedom that is attached to each sublattice (say, ``up'' on A and ``down'' on B), but within a given loop these pseudospins are flipped with probability 1/2. Then we accept an update if and only if the pseudospin of $\mu$ matches the pseudospin of $\nu$. If $\mu$ and $\nu$ belong to the same loop, then since they belong to the same sublattice their pseudospins must match and we always accept this update (in agreement with the probability list given above). However, if $\mu$ and $\nu$ belong to different loops, then their relative pseudospins are randomized and will only match 1/2 the time, and we end up accepting this type of move with probability 1/2 (again in agreement with the probability list given above).

\subsection{Updates in combined bond-spin space}

The fast updates described above can alternatively be formulated in the combined space of the valence bond basis and $S^z$ component spin basis. As discussed by Sandvik and Evertz~\cite{Sandvik-PRB-10}, this basis can be formed by considering the possible $S^z$ configurations on top of the valence bond configurations. Since all overlap graphs in the valence bond basis considered in this work form systems of closed loops, the only allowed $S^z$ configurations consist of alternating up and down spins within a loop. In general, the pattern of up and down spins between separate loops do not need to agree, and it is easy to see that this combined picture provides a physical motivation for the pseudospin degree of freedom introduced in the fast updates described above. Sandvik and Evertz only considered pure valence bond overlap graphs (i.e., they did not consider the type of internal triplet states we consider in this work for the calculation of the strange correlator), so here we show that overlaps between the internal triplet state and an arbitrary valence bond state produce the pseudospin pattern described above.

To derive the appropriate pseudospin pattern used in the fast updates, it is convenient to write the internal triplet state as a product state in the $S^{y}$ basis:
\begin{equation}
\ket{\Omega} = \prod_i \bigket{S_{i}^{y}=0}.
\end{equation}
For spin $S=1$, the on-site state $\ket{S_{i}^{y}=0}$ can be written in terms of the spin-1/2 degrees of freedom projected along the $z$ axis as
\begin{equation}
\sqrt{2}\bigket{S_{i}^{y}=0}=\ket{\uparrow\uparrow}+\ket{\downarrow\downarrow}
\end{equation}
so that the $AABB$ pattern within each closed loop will contain a pseudospin pattern of either $\uparrow\uparrow\downarrow\downarrow$ or $\downarrow\downarrow\uparrow\uparrow$. For higher spin, we simply form the symmetric combination of $S$ pairs in the $S_{i}^{y}=0$ state listed above. Thus, our internal triplet state can only be formed for {\it integral} $S$. In general, the loops formed in the overlap graph state between these internal triplet states and an arbitrary valence bond configuration will always obey an $AABB$ pattern, and the form of the internal triplet state described above allows for two possible spin arrangements in the $S^{z}$ basis: $\uparrow\uparrow\downarrow\downarrow$ and $\downarrow\downarrow\uparrow\uparrow$. Hence, the pseudospin update can also be applied to calculations of the strange correlator and represents a generalization of the fast updates in the combined bond-spin space described by Sandvik and Evertz~\cite{Sandvik-PRB-10}.

\section{\label{SEC:AppB}Rules for measuring spin-correlation functions}

The above rules for loop updates may also be used to construct rules for measuring spin correlation functions. They can be summarized by the following formula for spin flavors $\mu$ and $\nu$:
\begin{equation}
\vec{s}_{\mu} \cdot \vec{s}_{\nu} = \delta_{\mu \leftrightarrow \nu}\biggl( \frac{1}{2}\epsilon_{i_{\mu},j_{\nu}}+\frac{1}{4}\lambda_{\mu \leftrightarrow \nu} \biggr),
\end{equation}
where $\delta_{\mu \leftrightarrow \nu}$ is a $\delta$ function that triggers when $\mu$ and $\nu$ are in the same loop, while $i_{\mu}$ and $j_{\nu}$ are the sites hosting the spin flavors $\mu$ and $\nu$, respectively. As in the main text, $\epsilon_{i_{\mu},j_{\nu}}=\pm1$ for $i_{\mu}$ and $j_{\nu}$ on the same ($+1$) or opposite ($-1$) sublattices, and $\lambda_{\mu \leftrightarrow \nu}=\pm1$ for $\mu$ and $\nu$ with even ($+1$) or odd ($-1$) loop distance.

For valence bond overlaps in the singlet sector, the $ABAB$ pattern within each loop guarantees that when $i_{\mu}$ and $j_{\nu}$ belong to the same (opposite) sublattice, their loop spacing is even (odd). Thus, the above formula reduces to
\begin{equation}
\vec{s}_{\mu} \cdot \vec{s}_{\nu} = \frac{3}{4}\delta_{\mu \leftrightarrow \nu}\epsilon_{i_{\mu},j_{\nu}}.
\end{equation}
This is the standard formula for spin correlations in the valence bond basis. Beach and Sandvik have extended this to the case of four-spin correlations~\cite{Beach-NPB-06},
\begin{multline}
\bigl(\vec{s}_{\mu}\cdot\vec{s}_{\nu}\bigr)\bigl(\vec{s}_{\sigma}\cdot\vec{s}_{\tau}\bigr) = 
\epsilon_{i_{\mu},j_{\nu}}\epsilon_{k_{\sigma},l_{\tau}}\\
\times \biggl[\frac{3}{16}\bigr(\delta'_{\mu\leftrightarrow\nu}-\delta_{\mu\leftrightarrow\nu}\bigr)+\frac{9}{16}\delta'_{\mu\leftrightarrow\nu}\delta_{\sigma\leftrightarrow\tau}\biggr].
\end{multline}
Here, $\delta'_{\mu\leftrightarrow\nu}$ triggers only if $\mu$ and $\nu$ belong to the same loop {\it after} acting on the overlap graph with $\vec{s}_{\sigma}\cdot\vec{s}_{\tau}$. Since the operation of $\vec{s}_{\sigma}\cdot\vec{s}_{\tau}$ may merge, rearrange, or split existing loops, this is $\delta$ function is distinct from the unprimed one. For mixed overlap states, a more general relation can be derived:
\begin{equation}
\begin{split}
&\bigl(\vec{s}_{\mu}\cdot\vec{s}_{\nu}\bigr)\bigl(\vec{s}_{\sigma}\cdot\vec{s}_{\tau}\bigr) = \\
&\quad\frac{1}{16}\epsilon_{k_{\sigma},l_{\tau}}
\Bigl[
\bigl(2\epsilon_{i_{\mu},j_{\nu}}+\lambda'_{\mu\leftrightarrow\nu}\bigr)\delta'_{\mu\leftrightarrow\nu}\\
&\qquad\qquad\qquad\qquad\qquad-
\bigl(2\epsilon_{i_{\mu},j_{\nu}}+\lambda_{\mu\leftrightarrow\nu}\bigr)\delta_{\mu\leftrightarrow\nu}
\Bigr] \\
&+\frac{1}{16}\bigl(2\epsilon_{i_{\mu},j_{\nu}}+\lambda'_{\mu\leftrightarrow\nu}\bigr)\bigl(2\epsilon_{k_{\sigma},l_{\tau}}+\lambda_{\sigma\leftrightarrow\tau}\bigr)\delta'_{\mu\leftrightarrow\nu}\delta_{\sigma\leftrightarrow\tau}.
\end{split}
\end{equation}
Again, the prime indicates the values in the updated configuration obtained by acting $\vec{s}_{\sigma}\cdot\vec{s}_{\tau}$ on the overlap graph. This simplifies to the normal relation whenever $\lambda_{\mu\leftrightarrow\nu}=\epsilon_{i_{\mu},j_{\nu}}$.

In the next two subsections, we derive each component of the above expressions.

\subsection{Using singlet projection operator updates}

When the flavors $\mu$ and $\nu$ belong to opposite sublattices, we can decompose the spin correlations as $ \vec{s}_{\mu}\cdot\vec{s}_{\nu} = 1/4 - \hat{P}_{\mu,\nu}$. Since we already know the weights of $\hat{P}_{\mu,\nu}$, we can easily determine the weights of $ \vec{s}_{\mu}\cdot\vec{s}_{\nu}$:
\begin{itemize}
\item[(i)] If $\mu$ and $\nu$ belong to different loops, then $\vec{s}_{\mu}\cdot\vec{s}_{\nu} = 1/4 - (1/4) = 0$.
\item[(ii)] If $\mu$ and $\nu$ belong to the same loop with even loop spacing, then $\vec{s}_{\mu}\cdot\vec{s}_{\nu} = 1/4 - (1/2) = -1/4$.
\item[(iii)] If $\mu$ and $\nu$ belong to the same loop with odd loop spacing, then $\vec{s}_{\mu}\cdot\vec{s}_{\nu} = 1/4 - (1) = -3/4$.
\end{itemize}

\subsection{Using parity operator updates}

When the flavors $\mu$ and $\nu$ belong to the same sublattice, we instead decompose the spin correlations as $\vec{s}_{\mu}\cdot\vec{s}_{\nu} = \hat{Q}_{\mu,\nu}-1/4$. Since we already know the weights of $\hat{Q}_{\mu,\nu}$, we can easily determine the weights of $\vec{s}_{\mu}\cdot\vec{s}_{\nu}$:
\begin{itemize}
\item[(i)] If $\mu$ and $\nu$ belong to different loops, then $\vec{s}_{\mu}\cdot\vec{s}_{\nu} = 1/4 - 1/4 = 0$.
\item[(ii)] If $\mu$ and $\nu$ belong to the same loop with even loop spacing, then $\vec{s}_{\mu}\cdot\vec{s}_{\nu} = 1 - 1/4 = 3/4$.
\item[(iii)] If $\mu$ and $\nu$ belong to the same loop with odd loop spacing, then $\vec{s}_{\mu}\cdot\vec{s}_{\nu} = 1/2 - 1/4 = 1/4$.
\end{itemize}

\end{document}